\documentclass[format=acmsmall, review=false, screen=true, pdftex,hyperref,svgnames]{acmart}

\pdfoutput=1

\usepackage{booktabs} 
\usepackage{subcaption}
\usepackage{listings}
\usepackage{alltt}

\usepackage{tikz}
\usetikzlibrary{positioning}
\usetikzlibrary{matrix}
\usetikzlibrary{fit,calc}

\usepackage[ruled]{algorithm2e} 

\SetAlFnt{\small}
\SetAlCapFnt{\small}
\SetAlCapNameFnt{\small}
\SetAlCapHSkip{0pt}
\IncMargin{-\parindent}

\acmJournal{TOPS}
\acmVolume{9}
\acmNumber{4}
\acmArticle{39}
\acmYear{2010}
\acmMonth{3}
\copyrightyear{2009}

\setcopyright{acmlicensed}

\acmDOI{0000001.0000001}

\received{February 2007}
\received[revised]{March 2009}
\received[accepted]{June 2009}

\newif\ifsubmit\submitfalse
\ifsubmit

\newcommand{\mwh}[1]{}
\newcommand{\awr}[1]{}
\newcommand{\jp}[1]{}
\newcommand{\JP}[1]{}
\newcommand{\pxm}[1]{}
\newcommand{\michelle}[1]{}
\newcommand{\dml}[1]{}
\newcommand{\djv}[1]{}
\else

\newcommand{\mwh}[1]{\textcolor{blue}{MWH -- #1}}
\newcommand{\awr}[1]{\textcolor{green}{AWR -- #1}}
\newcommand{\jp}[1]{\textcolor{BlueViolet}{JP -- #1}}
\newcommand{\JP}[1]{\textcolor{BlueViolet}{JP -- #1}}
\newcommand{\pxm}[1]{\textcolor{orange}{PM -- #1}}
\newcommand{\michelle}[1]{\textcolor{teal}{MLM -- #1}}
\newcommand{\dml}[1]{\textcolor{purple}{dml -- #1}}
\newcommand{\djv}[1]{\textcolor{magenta}{DJV -- #1}}
\fi
\let\oldfootnote\footnote
\renewcommand{\footnote}[1]{\oldfootnote{\small #1}}
\newcommand{\bibifi}[0]{BIBIFI\xspace}
\newcommand{\originalmodel}[1]{}

\definecolor{insignificant}{rgb}{0.0,0.0,0.0}
\newcommand{\significant}[1]{\textbf{#1}}
\newcommand{\ahs}[0]{\hspace*{0.37em}}

\newcommand{\securelog}[0]{Secure Log\xspace}
\newcommand{\atm}[0]{ATM\xspace}
\newcommand{\ehr}[0]{Multiuser DB\xspace}

\definecolor{usercolor}{RGB}{47, 46, 51}
\definecolor{liocolor}{RGB}{10, 55, 104}
\definecolor{yesodcolor}{RGB}{11, 104, 51}
\definecolor{dbcolor}{RGB}{47, 46, 51}

\lstdefinestyle{customc}{
  belowcaptionskip=1\baselineskip,
  breaklines=true,
  numbers=left,
  xleftmargin=\parindent,
  language=C,
  columns=flexible,
  showstringspaces=false,
  basicstyle=\small\ttfamily,
  literate={{<-}{{$\leftarrow\,$}}2
            {->}{{$\rightarrow\,$}}2
            {<=}{{$\leq\,$}}2},
  numberstyle=\tiny\rmfamily,
}
\begin{document}
\title[Build It, Break It, Fix It]{Build It, Break It, Fix It: Contesting Secure Development}  

\author{James Parker}
\affiliation{
  \institution{University of Maryland, College Park}
  \streetaddress{8125 Paint Branch Dr}
  \city{College Park}
  \state{MD}
  \postcode{20740}
  \country{USA}}
\email{jp@jamesparker.me}

\author{Michael Hicks}
\affiliation{
  \institution{University of Maryland, College Park}
  \streetaddress{8125 Paint Branch Dr}
  \city{College Park}
  \state{MD}
  \postcode{20740}
  \country{USA}}
\email{mwh@cs.umd.edu}

\author{Andrew Ruef}
\affiliation{
  \institution{University of Maryland, College Park}
  \streetaddress{8125 Paint Branch Dr}
  \city{College Park}
  \state{MD}
  \postcode{20740}
  \country{USA}}
\email{awruef@cs.umd.edu}
%

\author{Michelle L. Mazurek}
\affiliation{
  \institution{University of Maryland, College Park}
  \streetaddress{8125 Paint Branch Dr}
  \city{College Park}
  \state{MD}
  \postcode{20740}
  \country{USA}}
\email{mmazurek@cs.umd.edu}

\author{Dave Levin}
\affiliation{
  \institution{University of Maryland, College Park}
  \streetaddress{8125 Paint Branch Dr}
  \city{College Park}
  \state{MD}
  \postcode{20740}
  \country{USA}}
\email{dml@cs.umd.edu}

\author{Daniel Votipka}
\affiliation{
  \institution{University of Maryland, College Park}
  \streetaddress{8125 Paint Branch Dr}
  \city{College Park}
  \state{MD}
  \postcode{20740}
  \country{USA}}
\email{dvotipka@cs.umd.edu}

\author{Piotr Mardziel}
\affiliation{
  \institution{Carnegie Mellon University}
  \streetaddress{4720 Forbes Avenue}
  \city{Pittsburgh}
  \state{PA}
  \postcode{15213}
  \country{USA}}
\email{piotrm@gmail.com}

\author{Kelsey R. Fulton}
\affiliation{
  \institution{University of Maryland, College Park}
  \streetaddress{8125 Paint Branch Dr}
  \city{College Park}
  \state{MD}
  \postcode{20740}
  \country{USA}}
\email{kfulton@cs.umd.edu}


\begin{abstract}
  Typical security contests focus on breaking or mitigating the impact
  of buggy systems. We present the Build-it, Break-it, Fix-it
  (\bibifi) contest, which aims to assess the ability to securely
  build software, not just break it. In \bibifi, teams build specified
  software with the goal of maximizing correctness, performance, and
  security. The latter is tested when teams attempt to break other
  teams' submissions. Winners are chosen from among the best builders
  and the best breakers. \bibifi was designed to be open-ended---teams
  can use any language, tool, process, etc. that they like. As such,
  contest outcomes shed light on factors that correlate with
  successfully building secure software and breaking insecure
  software. We ran three contests involving a total of 156 
  teams and three different programming problems. 
  Quantitative analysis
  from these contests found that the most efficient build-it
  submissions used C/C++, but submissions coded in a statically-type safe
  language were $11\times$ less likely to have a security flaw than C/C++ submissions.
	Break-it teams that were also successful build-it teams were
  significantly better at finding security bugs.
\end{abstract}

\ccsdesc[300]{Security and privacy~Software security engineering}
\ccsdesc[300]{Social and professional topics~Computing education}
\ccsdesc[300]{Software and its engineering~Software development techniques}

%
%

\keywords{Contest; Security; Education; Software; Engineering}

\maketitle

\renewcommand{\shortauthors}{J. Parker et al.}

\section{Introduction}
\label{sec:intro}
Cybersecurity
contests~\cite{mdc3,nationalccdc,defconctf,csawctf,cansec}
are popular proving grounds for cybersecurity
talent. Existing contests largely focus on \emph{breaking} (e.g.,
exploiting vulnerabilities or misconfigurations) and
\emph{mitigation} (e.g., rapid patching or reconfiguration). They do
not, however, test contestants' ability to \emph{build} (i.e., design
and implement) systems that are secure in the first place. Typical
programming contests~\cite{topcoder,acmprogramming,icfpcontest} do
focus on design and implementation, but generally ignore
security. This state of affairs is unfortunate because experts have
long advocated that achieving security in a computer system requires
treating security as a first-order design goal~\cite{SaltzerSc75}, and
is not something that can be added after the fact. As such, we should not
assume that good breakers will necessarily be good
builders~\cite{marketplace-defcon}, or that top coders can produce
secure systems.

This paper presents \textbf{Build-it, Break-it, Fix-it} (\bibifi),
a new security contest with a focus on \emph{building secure
  systems}.
A \bibifi contest has three phases. The first phase, \emph{Build-it},
asks small development teams to build software according to a provided
specification including security goals. The software is scored for
being correct, efficient, and featureful. The second phase,
\emph{Break-it}, asks teams to find defects in other teams' build-it
submissions. Reported defects, proved via test cases vetted by an
oracle implementation, benefit a
break-it team's score and penalize the build-it team's score; more
points are assigned to security-relevant problems. (A team's break-it
and build-it scores are independent, with prizes for top scorers in
each category.)  The final phase, \emph{Fix-it}, asks builders to fix
bugs and thereby get points back if the process discovers that distinct
break-it test cases identify the same defect.

\bibifi's design aims to
minimize the manual effort of running a contest, helping it scale. \bibifi's
structure and scoring system also aim to
encourage meaningful outcomes, e.g., to ensure that the top-scoring
build-it teams really produce secure and efficient software. Behaviors
that would thwart such outcomes are discouraged. For example, break-it
teams may submit a limited number of bug reports per build-it
submission, and will lose points during fix-it for test cases that
expose the same underlying defect or a defect also identified by other
teams. As such, they are encouraged to look for bugs broadly (in many
submissions) and deeply (to uncover hard-to-find bugs).

In addition to providing a novel educational experience, \bibifi
presents an opportunity to study the building and breaking process
scientifically. In particular, \bibifi contests may serve as a
quasi-controlled experiment that correlates participation data with
final outcome. By examining artifacts and participant surveys, we can
study how the choice of build-it programming language, team size and
experience, code size, testing technique, etc. can influence a team's
(non)success in the build-it or break-it phases. To the extent
that contest problems are realistic and contest participants represent
the professional developer community, the results of this study may
provide useful empirical evidence for practices that help or harm
real-world security. Indeed, the contest environment could be used to
incubate ideas to improve development security, with the best ideas
making their way to practice.

This paper studies the outcomes of three \bibifi contests that we held
during 2015 and 2016, involving three different programming problems. The first
contest asked participants to build a \emph{secure, append-only log}
for adding and querying events generated by a hypothetical art gallery
security system. Attackers with direct access to the log, but lacking
an ``authentication token,'' should not be able to steal or corrupt
the data it contains. The second contest 
asked participants to build a pair of
\emph{secure, communicating programs}, one representing an ATM and the
other representing a bank. Attackers acting as a man in the middle (MITM)
should neither be able to steal information (e.g., bank account names
or balances) nor corrupt it (e.g., stealing from or adding money to
accounts).\footnote{Such attacks are realistic in practice, as
detailed in a 2018 analysis of actual ATMs~\cite{atm-hacks}.}
The third contest required participants to build
a \emph{access-controlled, multiuser
data server} that protects the data of users. Users are authenticated via
password checking, and access control with delegation restricts how data is
read and modified. All contests drew participants from a MOOC
(Massive Online Open Courseware) course on cybersecurity. MOOC participants
had an average of 10 years of programming experience and had just completed a
four-course sequence including courses on secure software and cryptography.
The second and third contests also involved graduate and 
undergraduate students with less experience and training.
The first contest had 156 MOOC participants (comprising 68 teams). 
The second contest was composed of 122 MOOC participants (comprising 41 teams) and 
23 student participants (comprising 7 teams). 
The last contest had 68 MOOC participants (comprising 25 teams) and
37 student participants (comprising 15 teams).
%

\bibifi's design permitted it to scale reasonably well. For example,
one full-time person and two part-time judges ran the first 
contest in its entirety. 
This contest involved over one hundred participants 
who submitted more than 20,000 test cases. And
yet, organizer effort was limited to judging that the few hundred
submitted fixes addressed only a single conceptual defect; other work
was handled automatically or by the participants themselves. 

Rigorous quantitative analysis of the contests' outcomes revealed several
interesting, statistically significant effects. 
Considering build-it
scores: Writing code in C/C++ increased build-it scores initially, but
also increased chances of a security bug being found later; 
C/C++ programs were $11\times$ more likely to 
have a reported security bug
than programs written in a statically-type safe language. 
Considering break-it scores:
Larger teams found more bugs during the break-it phase. 
Break-it teams that also qualified during the build-it
phase found more security bugs than those
that did not. Use of advanced tools such as fuzzing or static analysis 
did not provide a significant advantage. 

Other trends emerged, but did not quite reach significance. 
Build-it teams
with more developers helped produce more secure implementations. 
Greater
programming experience
was also
helpful for breakers. 

We manually examined both build-it and break-it
artifacts. 
Successful build-it teams typically employed third-party
libraries---e.g., SSL, NaCL, and BouncyCastle---to implement
cryptographic operations and/or communications, which 
provided primitives such as
randomness, nonces, etc. 
Unsuccessful teams 
typically failed to employ cryptography, implemented it incorrectly,
used insufficient randomness, or failed to use
authentication. Break-it
teams found clever ways to exploit security problems;
some MITM implementations were quite sophisticated.

This paper extends a previously published
 conference paper~\cite{Ruef:2016:BBF:2976749.2978382} to include data from
 an additional contest run, as well as more details about our overall
 experience. Since publishing that paper we have made the \bibifi code and
infrastructure publicly available so that others may run their own
competitions. We, and those at other universities, including 
Universit\"at Paderborn,
University of Pennsylvania, 
Texas A\&M University, 
and Carnegie Mellon University, 
have already used the \bibifi infrastructure to run contests in a
classroom setting. 
More information, data, and opportunities to participate are available at
\url{https://builditbreakit.org} and the BIBIFI codebase
is at \url{https://github.com/plum-umd/bibifi-code}. 

In summary, this paper makes two main contributions. First, it
presents \bibifi, a security contest that encourages building, not
just breaking. Second, it presents a detailed description of three
\bibifi contests along with success and failure stories as well as a quantitative
analysis of the results. The paper is organized as follows.
We present the design of \bibifi in~\S\ref{sec:contest} and describe
specifics of the contests we ran in~\S\ref{sec:contests}.
%
We present success and failure stories of contestants
in~\S\ref{sec:stories} and a quantitative analysis of the data we collected
in~\S\ref{sec:analysis}. 
We review related work in~\S\ref{sec:related} and conclude
in~\S\ref{sec:conclusions}.

\section{Build-it, Break-it, Fix-it}
\label{sec:contest}
This section describes the goals, design, and implementation of the
\bibifi competition.
At the highest level, our aim is to create an environment that closely
reflects real-world development goals and constraints, and to encourage
build-it teams to write the most secure code they can, and break-it
teams to perform the most thorough, creative analysis of others' code
they can.
We achieve this through a careful design of how the competition is run
and how various acts are scored (or penalized). We also aim to
minimize the manual work required of the organizers---to allow the
contest to scale---by employing automation and proper participant
incentives.

\subsection{Competition phases} 
\label{sec:design}

We begin by describing the high-level mechanics of what occurs during a
\bibifi competition.
\bibifi may be administered on-line, rather than on-site, so teams may
be geographically distributed. The contest comprises three phases, each
of which last about two weeks for the contests we describe in this
paper.

\bibifi begins with the \textbf{build-it phase}.  Registered
contestants aim to implement the target software system according to a
published specification created by the contest organizers. A
suitable target is one that can be completed by good programmers in a
short time (just about two weeks, for the contests we ran), is easily
benchmarked for performance, and has an interesting attack surface. The
software should have specific security goals---e.g., protecting private
information or communications---which could be compromised by poor
design and/or implementation. The software should also not be too
similar to existing software to ensure that contestants do the
coding themselves (while still taking advantage of high-quality libraries and
frameworks to the extent possible). The software must build and run on a standard Linux VM made
available prior to the start of the contest.  Teams must develop using
Git~\cite{git}; with each push, the contest infrastructure downloads
the submission, builds it, tests it (for correctness and performance),
and updates the scoreboard. \S\ref{sec:contests} describes the
three target problems we developed: (1)~an append-only log (aka, \securelog), 
(2)~a pair of communicating programs that simulate a bank and ATM
(aka, \atm),
and (3)~a multi-user data server with custom access control policies
(aka, \ehr). 



The next phase is the \textbf{break-it phase}.
Break-it teams can download, build, and inspect all qualifying build-it
submissions, including source code; to qualify, the submission must
build properly, pass all correctness tests, and not be purposely
obfuscated (accusations of obfuscation are manually judged by the
contest organizers).  
We randomize each break-it team's
view of the build-it teams' submissions, but organize them by meta-data,
such as programming language used. (Randomization aims to encourage 
equal scrutiny of submissions by discouraging 
break-it teams from investigating projects in the same
order.) When they think they have found a defect, breakers submit a test case
that exposes the defect and an explanation of the issue.
We impose an upper bound on the number of test
cases a break-it team can submit against a single build-it submission,
to encourage teams to look at many submissions.
%
\bibifi's infrastructure automatically judges whether a submitted test
case truly reveals a defect. For example, for a correctness bug, it
will run the test against a reference implementation (``the oracle'')
and the targeted submission, and only if the test passes on the former
but fails on the latter will it be accepted. Teams
can also earn points by reporting bugs in the oracle, i.e., where its
behavior contradicts the written specification; these reports are
judged by the organizers.
More points are awarded to clear security problems, which may be
demonstrated using alternative test formats. The auto-judgment
approaches we developed for the three different contest problems are
described in~\S\ref{sec:contests}. 


The final phase is the \textbf{fix-it phase}.
Build-it teams are provided with the bug reports and test cases
implicating their submission. They may fix flaws these test cases
identify; if a single fix corrects more than one failing test case, the
test cases are ``morally the same,'' and thus points are only deducted
for one of them.  The organizers determine, based on information
provided by the build-it teams and other assessment, whether a
submitted fix is ``atomic'' in the sense that it corrects only one
conceptual flaw; if not, the fix is rejected.


Once the final phase concludes, prizes are awarded to the builders
and breakers with the best scores, as determined by the scoring
system described next.

\subsection{Competition scoring} 
\label{sec:contest-goals}

\bibifi's scoring system aims to encourage the contest's basic goals,
which are that the winners of the build-it phase truly
produced the highest quality software, and that the winners of the
break-it phase performed the most thorough, effective analysis of
others' code. The scoring rules, and the fact that scores are
published in real time while the contest takes place, create incentives
for good behavior (and disincentives for bad behavior). 

\subsubsection{Build-it scores} 
To reflect real-world development concerns, the winning build-it team
would ideally develop software that is correct, secure, and efficient.
While security is of primary interest to our contest,
developers in practice must balance these other aspects of quality
against security~\cite{cfi,ulfar-personal}, leading to a set of
trade-offs that cannot be ignored if we wish to motivate realistic
developer decision-making.

As such, each build-it team's score is the sum of the
\emph{ship} score\footnote{The name is meant to evoke a quality
  measure at the time software is shipped.} and the \emph{resilience} score.
The ship score is composed of points gained for correctness tests and
performance tests.  Each mandatory correctness test is worth $M$
points, for some constant $M$, while each optional correctness test is
worth $M/2$ points.
Each performance test has a numeric measure depending on the specific
nature of the programming project---e.g., latency, space consumed,
files left unprocessed---where lower measures are better.
A test's worth is $M \cdot{} (\mathit{worst} - v) / (\mathit{worst} -
\mathit{best})$, where $v$ is the measured result, $\mathit{best}$ is
the measure for the best-performing submission, and $\mathit{worst}$ is
the worst performing. As such, each performance test's value ranges
from 0 to $M$. As scores are published in real time, teams can see
whether they are scoring better than other participants. Their
relative standing may motivate them to improve their
implementation to improve its score before the build-it phase ends.

The resilience score is determined after the break-it and fix-it
phases, at which point the set of unique defects against a submission
is known. For each \emph{unique} bug found against a team's submission
we subtract $P$ points from its resilience score; as such, the best
possible resilience score is $0$. For correctness bugs, we set $P$ to
$M/2$; for crashes that violate memory safety we set $P$ to $M$, and
for exploits and other security property failures we set $P$ to
$2M$. (We discuss the rationale for these choices below.) Once again,
real-time scoring helps incentivize fixing, to get points back.

%

\subsubsection{Break-it scores} 
Our primary goal with break-it teams is to encourage them to find as
many defects as possible in the submitted software, as this would give
greater confidence in our assessment that one build-it team's
software is of higher quality than another's.  While we are
particularly interested in obvious security defects, correctness
defects are also important, as they can have non-obvious security
implications.

A break-it team's score is the summed value
of all defects they have found, using the above $P$ valuations. This
score is shown in real time during the break-it phase, incentivizing teams to
improve their standing. After
the fix-it phase, this score is reduced. In particular, if a break-it team
submitted multiple test cases against a project that identify the same
defect, the duplicates are discounted. Moreover, each of the
$N$ break-it teams' scores that identified the same defect are adjusted
to receive $P/N$ points for that defect, splitting the $P$ points among
them.  

Through a combination of requiring concrete test cases and scoring,
\bibifi encourages break-it teams to follow the spirit of the
competition.
First, by requiring them to provide test cases as evidence of a defect
or vulnerability, we ensure they are providing useful bug reports.
By providing $4\times$ more points for security-relevant bugs than for
correctness bugs, we
nudge break-it teams to look for these sorts of flaws, and to not just
focus on correctness issues. (But a different ratio might work better;
see below.) Because break-it teams are limited to a fixed number of test cases per
submission, they are discouraged from submitting many tests they
suspect are ``morally the same;'' as they could lose points
for them during the fix-it phase they are better off submitting
tests demonstrating different bugs.
Limiting per-submission test cases also encourages examining many
submissions. Finally, because points for defects found by other teams
are shared, break-it teams are encouraged to look for hard-to-find bugs, rather than
just low-hanging fruit.

\subsubsection{Discouraging collusion} 
\bibifi contestants may form teams however they wish, and may
participate remotely.
This encourages wider participation, but it also opens the possibility
of collusion between teams, as there cannot be a judge overseeing their
communication and coordination.
There are three broad possibilities for collusion, each of which 
\bibifi's scoring discourages.

First, two break-it teams could consider sharing bugs they find with
one another.  By scaling the points each finder of a particular bug
obtains, we remove incentive for them to both submit the same bugs, as
they would risk diluting how many points they both obtain.

The second class of collusion is between a build-it team and a
break-it team, but neither have incentive to assist one another.
The zero-sum nature of the scoring between breakers and builders
places them at odds with one another; revealing a bug to a break-it
team hurts the builder, and not reporting a bug hurts the breaker.

Finally, two build-it teams could collude, for instance by sharing code
with one another. It might be in their interests to do this in the
event that the competition offers prizes to two or more build-it teams,
since collusion could obtain more than one prize-position.  We use
judging and automated tools (and feedback from break-it teams) to
detect if two teams share the same code (and disqualify them), but it
is not clear how to detect whether two teams provided out-of-band
feedback to one another prior to submitting code (e.g., by holding
their own informal ``break-it'' and ``fix-it'' stages).  We view this
as a minor threat to validity; at the surface, such assistance appears
unfair, but it is not clear that it is contrary to the goals of the
contest, that is, to developing secure code.

\subsection{Discussion}

The contest's design also aims to enable scalability by reducing work
on contest organizers. In our experience, \bibifi's design succeeds at
what it sets out to achieve, but has limitations.

\paragraph*{Minimizing manual effort}
Once the contest begins, manual effort by the organizers is limited by
design. All bug reports submitted during the break-it phase are
automatically judged by the oracle; organizers only need to vet any
bug reports against the oracle itself. Organizers may also need to judge
accusations by breakers of code obfuscation by builders. Finally,
organizers must judge whether submitted fixes address a single defect;
this is the most time consuming task. It is necessary because we
cannot automatically determine whether multiple bug reports against
one team map to the same software defect; techniques for
automatic testcase deduplication are still a matter of research (see
\S\ref{sec:related}). As such, we incentivize
build-it teams to demonstrate overlap through fixes, which organizers
manually confirm address only a single defect, not several.

Previewing some of the results presented later, we can confirm that
the design works reasonably well. For example, as detailed in
Table~\ref{tab:bugs-fixes}, 68 teams submitted 24,796 test cases for
the \securelog contest. The oracle auto-rejected 15,314 of these, and
build-it teams addressed 2,252 of those remaining with 375 fixes, a
6$\times$ reduction. Most confirmations that a fix truly addresses a
single bug took 1-2 minutes each. Only 30 of these fixes were
rejected. No accusations of code obfuscation were made by break-it
teams, and few bug reports were submitted against the oracle. All
told, the \securelog contest was successfully managed by one
full-time person, with two others helping with judging.

\paragraph*{Limitations}
While we believe \bibifi's structural and scoring incentives are
properly designed, we should emphasize several limitations. 

First, there is no guarantee that all implementation defects will be
found. Break-it teams may lack the time or skill to find problems in
all submissions, and not all submissions may receive equal
scrutiny. Break-it teams may also act contrary to incentives and focus
on easy-to-find and/or duplicated bugs, rather than the harder and/or
unique ones. 
In addition, certain vulnerabilities, like insufficient randomness in key generation, may take more 
effort to exploit, so breakers may skip such vulnerabilities. 
Finally, break-it teams may find defects that the \bibifi
infrastructure cannot automatically validate, meaning those defects
will go unreported. However, with a large enough pool of break-it
teams, and sufficiently general defect validations automation, we still
anticipate good coverage both in breadth and depth.

Second, builders may fail to fix bugs in a manner that is in their
best interests. For example, in not wanting to have a fix rejected as
addressing more than one conceptual defect, teams may use several
specific fixes when a more general fix would have been allowed. 
Additionally, teams that are out of contention for prizes may
simply not participate in the fix-it phase.\footnote{Hiding scores
  during the contest might help mitigate this, but would harm incentives during break-it
  to go after submissions with no bugs reported against them.}
 We observed these behaviors in our contests. Both
actions decrease a team's resilience score (and correspondingly
increase breakers' scores). 
For our most recent contest, we attempted to create an incentive to fix bugs
by offering prizes to participants that scale with their final score, rather than offering 
prizes only to winners.
Unfortunately, this change in prize structure did not increase fix-it participation. 
We discuss fix-it behavior in more depth in~\S\ref{sec:resilience}.

Finally, there are several design points in a problem's definition and
testing code
that may skew results. For example, too few correctness tests may
leave too many correctness bugs to be found during break-it,
distracting break-it teams' attention from security issues. Too many
correctness tests may leave too few bugs, meaning teams are differentiated
insufficiently by general bug-finding ability. Scoring prioritizes
security problems 4 to 1 over correctness problems, but it is hard to
say what ratio makes the most sense when trying to maximize real-world
outcomes; both higher and lower ratios could be argued. How security
bugs are classified will also affect behavior; two of our contests had
strong limits on the number of possible security bugs per project,
while the third's definition was far more (in fact, too) liberal, as
discussed in \S\ref{sec:security-bug}.
Finally, performance
tests may fail to expose important design tradeoffs (e.g., space
vs. time), affecting the ways that teams approach maximizing their 
ship scores. For the contests we report in this paper, we are
fairly comfortable with these design points. In particular, our
pilot contest~\cite{bibifi-cset15} prioritized security bugs 2 to 1
and had fewer interesting performance tests, and outcomes were better
when we increased the ratio.

\subsection{Implementation} 
\label{sec:implementation}

\begin{figure}[t!]
  \begin{centering}
	\includegraphics[width=\columnwidth]{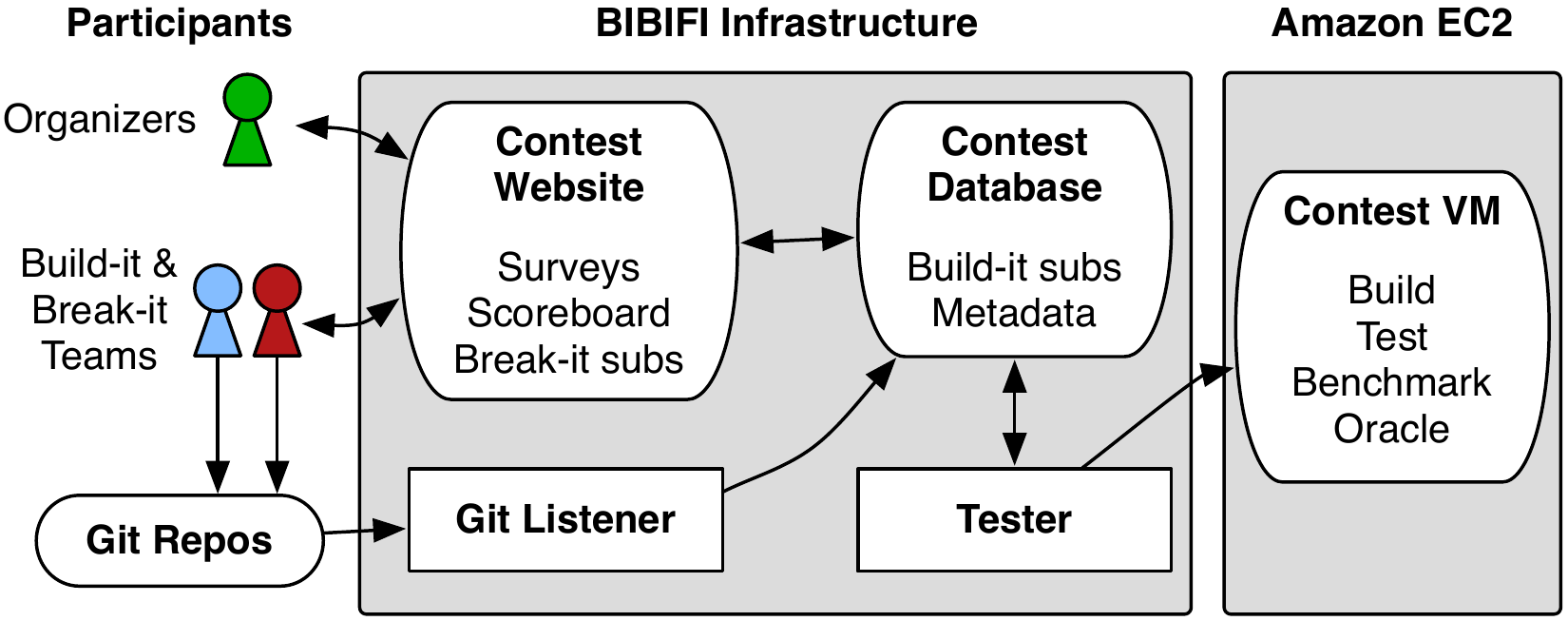}
  \end{centering}
  \caption{\label{fig:system} Overview of \bibifi's implementation.}
\end{figure}



Figure~\ref{fig:system} provides an overview of the \bibifi
implementation. It consists of a web frontend, providing the interface
to both participants and organizers, and a backend for testing builds
and breaks. Key goals of the infrastructure are security---we do not
want participants to succeed by hacking \bibifi itself---and
scalability. 

\paragraph*{Web frontend} 

%
Contestants sign up for the contest through our web application
frontend, and fill out a survey when doing so, to gather demographic
data potentially relevant to the contest outcome (e.g.,
programming experience and security training). During the contest, the
web application tests build-it submissions and break-it bug reports,
keeps the current scores updated, and provides a workbench for the
judges for considering whether or not a submitted fix covers one bug or
not.

To secure the web application against unscrupulous participants, we
implemented it in ${\sim}$11500 lines of Haskell using the Yesod
\cite{yesodweb} web framework backed by a PostgreSQL \cite{psql}
database. Haskell's strong type system defends against use-after-free,
buffer overrun, and other memory safety-based attacks.  The use of
Yesod adds further automatic protection against various attacks like
CSRF, XSS, and SQL injection. As one further layer of defense, the web
application incorporates the information flow control framework
LWeb, which is derived from LIO~\cite{lio},
in order to protect against inadvertent information leaks and privilege
escalations. LWeb dynamically guarantees that users can only access
their own information, as established by a mechanized proof of
correctness (in Liquid Haskell~\cite{Parker:2019:LIF:3302515.3290388}). 



\paragraph*{Testing backend} 

The backend infrastructure is used for
testing during the build-it phase for correctness and
performance, and during the break-it phase to assess potential
vulnerabilities. It consists of ${\sim}$5500 lines of
Haskell code (and a little Python). 

To automate testing, we require contestants to specify a URL to a
Git~\cite{git} repository (hosted on either Gitlab, Github, or Bitbucket) and
shared with a designated \texttt{bibifi} username, read-only. The
backend ``listens'' to each contestant repository for pushes, upon
which it downloads and archives each
commit. Testing is then handled by a scheduler that spins up a (Docker or Amazon
EC2) virtual machine which builds and tests each submission. We require
that teams' code builds and runs, without any network access, in an
Ubuntu Linux VM that we share in advance. Teams can request that we
install additional packages not present on the VM\@. The use of VMs
supports both scalability (Amazon EC2 instances are dynamically
provisioned) and security---using fresh VM instances prevents a team
from affecting the results of future tests, or of tests on other teams'
submissions. 

All qualifying build-it submissions may be downloaded by break-it
teams at the start of the break-it phase.  
%
%
As break-it teams identify bugs, they prepare a (JSON-based) file
specifying the buggy submission along with a sequence of commands with
expected outputs that demonstrate the bug. Break-it teams commit and push this
file (to their Git repository). The backend uses the file
to set up a test of the implicated submission to see if it indeed is a
bug. 

The code that 
tests build and break submissions differs for each contest problem. 
To increase modularity, we have created a problem API so that testing code run on the VM 
can easily be swapped out for different contest problems. 
Contest organizers can create their own problems by conforming to this API. 
The infrastructure will set up the VM and provide submission information to the problem's testing software via JSON. 
The problem's software runs the submission and outputs the result as JSON, which the infrastructure records and uses to update scores accordingly. 
Details are available in the documentation of the contest repository.

\section{Contest Problems} 
\label{sec:contests}
This section presents the three programming problems we have developed
for BIBIFI contests. These three problems were used during open
competitions in 2015 and 2016, and in our own and others'
undergraduate security courses since then. We discuss each 
problem and its specific notions of
security defect, as well as how breaks exploiting such defects are
automatically judged.
%

\subsection{\securelog}
\label{sec:problem}

The \securelog problem was motivated as support for an art gallery
security system. Contestants write two programs. The first,
\texttt{logappend}, appends events to the log; these events indicate
when employees and visitors enter and exit gallery rooms. The second,
\texttt{logread}, queries the log about past events. To qualify,
submissions must implement two basic queries (involving the current
state of the gallery and the movements of particular individuals), but they
could implement two more for extra points (involving time spent in the
museum, and intersections among different individuals' histories). An
empty log is created by \texttt{logappend} with a given authentication
token, and later calls to \texttt{logappend} and
\texttt{logread} on the same log must use that token or the
requests will be denied.

Here is a basic example of invocations to \texttt{logappend}. The
first command creates a log file called \texttt{logfile} because one
does not yet exist, and protects it with the authentication token 
\texttt{secret}. In addition, it records that \texttt{Fred}
entered the art gallery.  Subsequent executions of
\texttt{logappend} record the events of \texttt{Jill} entering the gallery and
both guests entering room \texttt{1}. 
\begin{lstlisting}
$ ./logappend -K secret -A -G Fred logfile
$ ./logappend -K secret -A -G Jill logfile
$ ./logappend -K secret -A -G Fred -R 1 logfile
$ ./logappend -K secret -A -G Jill -R 1 logfile
\end{lstlisting}
Here is an example of \texttt{logread}, using the logfile just
created. It queries who is in the gallery and what rooms they are currently in. 
\begin{lstlisting}
$ ./logread -K secret -S logfile
Fred
Jill
1: Fred,Jill
\end{lstlisting} 

The problem states that an attacker is allowed direct access to the
logfile, and yet integrity and privacy must be maintained. 
A canonical way of implementing the log is therefore to treat the
authentication token as a symmetric key for authenticated
encryption, e.g., using a combination of AES and HMAC. 
%
There are several tempting shortcuts that we anticipated build-it
teams would take (and that break-it teams would exploit).
For instance, one may be tempted to encrypt and sign individual log records
as opposed to the entire log, thereby making \texttt{logappend} faster.
But this could permit integrity breaks that duplicate or reorder log
records.
Teams may also be tempted to implement their own
encryption rather than use existing libraries, or to simply sidestep
encryption altogether.
\S\ref{sec:stories} reports several cases we observed.


A submission's performance is measured in terms of time to perform a
particular sequence of operations, and space consumed by the resulting
log. Correctness (and \emph{crash}) bug reports are defined as sequences of
\texttt{logread} and/or \texttt{logappend} operations with expected
outputs (vetted by the oracle).
Security is defined by \emph{privacy} and \emph{integrity}: any attempt to learn something
about the log's contents, or to change them, without the using
\texttt{logread} and \texttt{logappend} and the proper token should be
disallowed. How violations of these properties are specified and tested
is described next.



\paragraph*{Privacy breaks}
%
%
%
%
When providing a build-it submission to the break-it teams, we also
included a set of log files that were generated using a sequence of
invocations of that submission's \texttt{logappend} program. We
generated different logs for different build-it submissions, using a
distinct command sequence and authentication token for each. All logs
were distributed to break-it teams without the authentication token;
some were distributed without revealing the sequence of commands (the
``transcript'') that generated them. For these, a break-it team could
submit a test case involving a call to \texttt{logread} (with the
authentication token omitted) that queries the file. The \bibifi
infrastructure would run the query on the specified file with the
authentication token, and if the output matched that specified by the
breaker, then a privacy violation is confirmed.
%
For example, before the break-it round, the infrastructure would run a bunch of randomly generated commands against a given team's implementation.
\begin{lstlisting}
$ ./logappend -K secret -A -G Fred logfile
$ ./logappend -K secret -A -G Fred -R 816706605 logfile
\end{lstlisting} 
Breakers are only given the \texttt{logfile} and not the secret token or the transcript of the commands (for privacy breaks). 
A breaker would demonstrate a privacy violation by submitting the following string, which matches the invocation of \texttt{./logread -K secret -S logfile}.
\begin{lstlisting}
Fred
816706605: Fred
\end{lstlisting}
The system knows the breaker has 
successfully broken privacy since the breaker is able to present confidential information without knowing the secret token.
In practice, the transcript of commands is significantly longer and a random secret is used. 

\paragraph*{Integrity breaks}
For about half of the generated log files we also provided the
transcript of the \texttt{logappend} operations used to generate the
file. A team could submit a test case specifying the name of the log file, the
contents of a corrupted version of that file, and a \texttt{logread}
query over it (without the authentication token). For both the
specified log file and the corrupted one, the \bibifi infrastructure
would run the query using the correct authentication token. An
integrity violation is detected if the query command produces a
non-error answer for the corrupted log that 
differs from the correct answer (which can be confirmed against the
transcript using the oracle).


This approach to determining privacy and integrity breaks has the
benefit and drawback that it does not reveal the \emph{source} of the
issue, only that there is (at least) one, and that it is exploitable. As such, we only count up to
one integrity break and one privacy break against the score of
each build-it submission, even if there are multiple defects that could
be exploited to produce privacy/integrity violations (since we could
not automatically tell them apart).

\subsection{\atm}
\label{sec:atm-problem}

The \atm problem asks builders to construct two communicating programs:
\texttt{atm} acts as an ATM client, allowing customers to set up an
account, and deposit and withdraw money; \texttt{bank} is a
server that tracks client bank balances and processes their requests,
received via TCP/IP\@. \texttt{atm} and \texttt{bank} should
only permit a customer with a correct \emph{card file} to learn or
modify the balance of their account, and only in an appropriate way
(e.g., they may not withdraw more money than they have). In addition,
\texttt{atm} and \texttt{bank} should only communicate if they can
authenticate each other. They can use an \emph{auth file} for this
purpose; it will be shared between the two via a
trusted channel unavailable to the attacker.\footnote{In a real
deployment, this might be done by ``burning'' the auth file into the ATM's
ROM prior to installing it.}  Since the \texttt{atm} is communicating
with \texttt{bank} over the network, a ``man in the middle'' (MITM)
could observe and modify exchanged messages, or insert new messages.
The MITM could try to compromise security despite not having access to
auth or card files. Such compromise scenarios are realistic, even in
2018~\cite{atm-hacks}. 

Here is an example run of the \texttt{bank} and \texttt{atm} programs.
\begin{lstlisting}
$ ./bank -s bank.auth &
\end{lstlisting}
This invocation starts the \texttt{bank} server, which creates the
file \texttt{bank.auth}. This file will be used by the \texttt{atm}
client to authenticate the \texttt{bank}. The \texttt{atm} is started
as follows:
\begin{lstlisting}
$ ./atm -s bank.auth -c bob.card -a bob -n 1000.00
{"account":"bob","initial_balance":1000}
\end{lstlisting}
The client initiates creation of a new account for user \texttt{bob} with an initial
balance of \$1,000. It also creates a file \texttt{bob.card} that is
used to authenticate \texttt{bob} (this is basically Bob's PIN) from here on.  
A receipt of the transaction from the server is printed as JSON. The \texttt{atm}
client can now use \texttt{bob}'s card to perform further actions on
his account. For example, this command withdraws \$63.10 from
\texttt{bob}'s account: 
\begin{lstlisting}
$ ./atm -s bank.auth -c bob.card -a bob -w 63.10
{"account":"bob","withdraw":63.1}
\end{lstlisting} 

A canonical way of implementing the \texttt{atm} and \texttt{bank} 
programs would be to use public key-based authenticated and
encrypted communications. The auth file is used as the \texttt{bank}'s public
key to ensure that key negotiation initiated by the \texttt{atm} is
with the \texttt{bank} and not a MITM. When creating an account, the
card file should be a suitably large random number, so that the MITM is
unable to feasibly predict it. It is also necessary to protect against
replay attacks by using nonces or similar mechanisms.
As with \securelog, a wise approach
would be use a library like OpenSSL to implement these features. Both
good and bad implementations are discussed further in~\S\ref{sec:stories}. 

Build-it submissions' performance is measured as the time to complete
a series of benchmarks involving various \texttt{atm}/\texttt{bank} interactions.\footnote{The
transcript of interactions is always serial, so there was no motivation to
support parallelism for higher throughput.} Correctness 
(and \emph{crash}) bug reports are defined as sequences of
\texttt{atm} commands where the targeted submission produces
different outputs than the oracle (or crashes).
Security defects are specified as follows.



\paragraph*{Integrity breaks} 
Integrity violations are demonstrated using a custom MITM program
that acts as a proxy: It listens on a specified IP address and TCP
port, and accepts a connection from the \texttt{atm} while connecting to the
\texttt{bank}. We provided a Python-based proxy as a
starter MITM; it forwards communications between the endpoints after
establishing the connection. A breaker's MITM would modify this
baseline behavior, observing and/or modifying
messages sent between \texttt{atm} and \texttt{bank}, and perhaps
dropping messages or initiating its own. 

To demonstrate an integrity violation, the MITM will send requests to a
\emph{command server}. It can tell the
server to run inputs on the \texttt{atm} and it can ask for the
card file for any account whose creation it initiated. Eventually the MITM
will declare the test complete. At this point, the same set of
\texttt{atm} commands is run using the oracle's \texttt{atm} and
\texttt{bank} \emph{without the MITM}. This means that any messages
that the MITM sends directly to the target submission's \texttt{atm} or \texttt{bank}
will not be replayed/sent to the oracle.
If the oracle and target both
complete the command list without error, but they differ on the
outputs of one or more commands, or on the balances of
accounts at the bank whose card files were not revealed to the MITM
during the test, then there is evidence of an integrity violation.

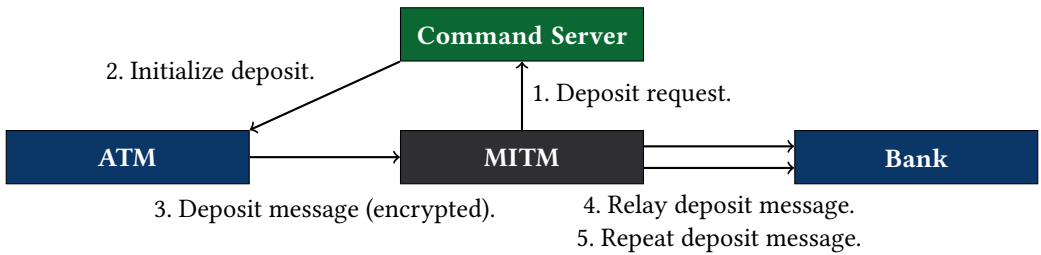
\begin{figure}
\begin{tikzpicture}[
mynode/.style={
  draw,
  text width=3cm,
  minimum height=0.7cm,
  align=center
  },
]
\node[mynode,fill=yesodcolor] (commandserver) {\textcolor{white}{\textbf{Command Server}}};

\node[mynode,fill=usercolor,below=0.9cm of commandserver] (mitm) {\textcolor{white}{\textbf{MITM}}};

\node[mynode,fill=liocolor,left=2cm of mitm] (atm) {\textcolor{white}{\textbf{ATM}}};
\node[mynode,fill=liocolor,right=2cm of mitm] (bank) {\textcolor{white}{\textbf{Bank}}};

\draw[<-,thick]
  (commandserver) -- node[right] {1. Deposit request.} (mitm);
\draw[<-,thick]
  (atm.north east) -- node[above left] {2. Initialize deposit.} (commandserver.south west);
\draw[<-,thick]
  (mitm) -- node[below=0.4cm] {3. Deposit message (encrypted).} (atm);
\draw[<-,thick]
  (bank.175) -- node[below=0.5cm] {4. Relay deposit message.} (mitm.5);
\draw[<-,thick]
  (bank.185) -- node[below=0.65cm] {5. Repeat deposit message.} (mitm.355);

%
%
%
%
\end{tikzpicture}
\caption{MITM replay attack.}
\label{fig:mitmattack}
\end{figure}

As an example (based on a real attack we observed), consider a
submission that uses deterministic encryption without nonces in
messages. The MITM could direct the command server to deposit money
from an account, and then replay the message it observes. When run on
the vulnerable submission, this would credit the account twice. But
when run on the oracle without the MITM, no message is replayed,
leading to differing final account balances. A correct submission
would reject the replayed message, which would invalidate the break.
This example is illustrated in Figure \ref{fig:mitmattack}.

\paragraph*{Privacy breaks} 
Privacy violations are also demonstrated using a MITM\@. In this case, at
the start of a test the command server will generate two random
values, \emph{amount} and \emph{account name.} If by the end of the test no
errors have occurred and the attacker can prove it knows the actual
value of either secret (by sending a command that specifies it),
the break is considered successful. Before
demonstrating knowledge of the secret, the MITM can
send commands to the and server with a \emph{symbolic} reference to \emph{amount}
and \emph{account name}; the server will fill in the
actual secrets before forwarding these messages. The command server does
not automatically create a secret account or an account with a secret
balance; it is up to the breaker to do that (referencing the secrets
symbolically when doing so). 

As an example, suppose the target does not encrypt exchanged messages. Then a
privacy attack might be for the MITM to direct the command server to
create an account whose balance contains the secret amount. Then the
MITM can observe an unencrypted message sent from \texttt{atm} to
\texttt{bank}; this message will contain the actual amount filled in
by the command server. The MITM can then send its guess to the command
server showing that it knows the amount. 

As with the \securelog problem, we cannot tell whether an integrity or
privacy test is exploiting the same underlying weakness in a
submission, so we only accept one violation of each category against
each submission.  

\paragraph*{Timeouts and denial of service}
One difficulty with our use of a breaker-provided MITM is that we cannot reliably detect
bugs in \texttt{atm} or \texttt{bank} implementations that would result in infinite loops, missed
messages, or corrupted messages. This is because such bugs can be
simulated by the MITM by dropping or corrupting messages it
receives. Since the builders are free to implement any protocol they
like, our auto-testing infrastructure cannot tell if a protocol
error or timeout is due to a bug in the target or due to misbehavior
of the MITM\@. As such, we conservatively disallow any MITM test run
that results in the target \texttt{atm} or \texttt{bank} hanging
(timing out) or returning with a protocol error (e.g., due to a corrupted packet). 
 This means that flaws in
builder implementations might exist 
but evidence of those bugs might not be realizable in our testing system. 






\subsection{\ehr}
\label{sec:ehr}

\begin{figure}
  \centering
  \begin{minipage}{.6\textwidth}
\begin{small}
\begin{alltt}
<prog>      ::= as principal {\it p} password {\it s} do \verb+\+n <cmd> ***
<cmd>        ::= exit \verb+\+n | return <expr> \verb+\+n | <prim_cmd> \verb+\+n <cmd>
<expr>     ::=  <value> | [] | { <fieldvals> }
<fieldvals> ::=  {\it x} = <value> | {\it x} = <value> , <fieldvals>
<value>    ::=  {\it x} | {\it x} . {\it y} | {\it s}
<prim_cmd> ::= create principal {\it p} s
            | change password {\it p} s
            | set {\it x}  = <expr>
            | append to {\it x} with <expr>
            | local {\it x} = <expr>
            | foreach {\it y} in {\it x} replacewith <expr>
            | set delegation <tgt> {\it q} <right> -> {\it p} 
            | delete delegation <tgt> {\it q} <right> -> {\it p} 
            | default delegator = {\it p} 
<tgt>      ::= all | {\it x}
<right>    ::= read | write | append | delegate
\end{alltt}
\end{small}
\end{minipage}
\caption{Grammar for the \ehr command language as BNF\@. Here, $x$ and $y$
  represent arbitrary variables; $p$ and $q$ represent arbitrary
  principals; and $s$ represents an arbitrary string. Commands
  submitted to the server should match the non-terminal \texttt{<prog>}.}
\label{fig:ehrgrammar}
\end{figure}

The \ehr problem requires builders to implement a server that maintains a
multi-user key-value store, where users' data is protected by
customizable access control policies.  The data server accepts queries
written in a text-based command language delivered over TCP/IP (we assume
the communication channel is trusted, for simplicity).  Each program
begins by indicating the querying user, authenticated with a
password. It then runs a sequence of commands to read and write data
stored by the server, where data values can be strings, lists, or records. 
The full grammar of the command language is shown in
Figure~\ref{fig:ehrgrammar} where the start symbol (corresponding to a
client command) is \lstinline{<prog>}. Accesses to data may be denied if the
authenticated user lacks the necessary permission. A user can delegate
permissions, like reading and writing variables, to other principals.
If running the command program results in a security violations or
error then all of its effects will be rolled back.

Here is an example run of the data server. To start, the server is
launched and listens on TCP port 1024. 
\begin{lstlisting}
$ ./server 1024 &
\end{lstlisting}
Next, the client submits the following program. 
\begin{lstlisting}
as principal admin password "admin" do
   create principal alice "alices_password"
   set msg = "Hi Alice. Good luck!"
   set delegation msg admin read -> alice
   return "success"
***
\end{lstlisting}
The program starts by declaring it is running on behalf of
principlal \lstinline{admin}, whose password is \lstinline{"admin"}. If
authentication is successful, the program creates a new principal
\texttt{alice}, creates a new variable \texttt{msg} containing a
string, and then delegates read permission on the variable to
Alice. The server sends back a transcript of the successful commands
to the client, in JSON format:
\begin{lstlisting}
{"status":"CREATE_PRINCIPAL"}
{"status":"SET"}
{"status":"SET_DELEGATION"}
{"status":"RETURNING","output":"success"}
\end{lstlisting}
Next, suppose Alice sends the following program, which simply logs in
and reads the \lstinline{msg} variable:
\begin{lstlisting}
as principal alice password "alices_password" do
   return msg
***
\end{lstlisting}
The server response indicates the result:
\begin{lstlisting}
{"status":"RETURNING","output":"Hi Alice. Good luck!"}
\end{lstlisting}

The data server is implemented by writing a parser to read the input command programs. 
The server needs a store to keep the value 
of each variable as well as an access control list that tracks the permissions for the variables. 
It also needs to keep track of delegated permissions, which can form
chains; e.g., Alice could delegate read permission to all of her
variables to Bob, who could then delegate permission to read one of
those variables to Charlie. If when executing the program a security
violation or other error occurs (e.g., reading a variable that doesn't
exist), the server needs to
roll back its state to what it was prior to processing the input program. 
All responses back to the client are encoded as JSON. 

\paragraph*{Scoring}
A data server's performance is measured in elapsed runtime to process sequences of programs. 
Correctness (and \emph{crash}) violations are demonstrated by providing a sequence of command programs where the data server's output differs from that of the oracle 
(or the implementation crashes). 
Security violations can be to data privacy, integrity, or
availability, by comparing the behavior of the target against that of
an oracle implementation.

\paragraph*{Privacy breaks}
A privacy violation occurs when the oracle would deny a request to
read a variable, but the target implementation allows it.
Consider the following example where a variable, \texttt{secret}, is created, but Bob is not allowed to read it.
\begin{lstlisting}
as principal admin password "admin" do
   create principal bob "bobs_password"
   set secret = "Super secret"
   return "success"
***

{"status":"CREATE_PRINCIPAL"}
{"status":"SET"}
{"status":"RETURNING","output":"success"}
\end{lstlisting}
Now Bob attempts to read the \texttt{secret} variable with the following query.
\begin{lstlisting}
as principal bob password "bobs_password" do
   return secret
***
\end{lstlisting}
Bob does not have permission to read \texttt{secret}, so the oracle returns \texttt{\{"status":"DENIED"\}}. 
If the implementation returns the \texttt{secret} contents of \texttt{\{"status":"RETURNING","output":"Super secret"\}}, we know a confidentiality violation has occurred. 

\paragraph*{Integrity breaks}
Integrity violations are demonstrated in a similar manner, but occur when unprivileged users modify variables they don't have permission to write to.
With the example above, the variable \texttt{secret}, is created, but Bob is not allowed to write to it. 
Now Bob attempts to write to \texttt{secret} with the following query.
\begin{lstlisting}
as principal bob password "bobs_password" do
   set secret = "Bob's grade is an A!"
   return "success"
***
\end{lstlisting}
Bob does not have write permission on \texttt{secret}, so the oracle returns \texttt{\{"status":"DENIED"\}}.
If the implementation returns the following, an integrity violation has been demonstrated.
\begin{lstlisting}
{"status":"SET"}
{"status":"RETURNING","output":"success"}
\end{lstlisting}

\paragraph*{Availability breaks}
Unlike the \atm problem, we are able to assess availability violations
for \ehr (since we are not using a MITM). In this case, a security violation is possible 
when the server implementation is unable to process legal command programs. 
This is demonstrated when the target incorrectly denies a program by reporting an
error, but the oracle successfully executes the program. 
Availability security violations also happen when the server
implementation fails to respond to an input program within a fixed period of time. 

\bigskip

Unlike the other two problems, the \ehr problem does not place a limit
on the number of security breaks submitted. 
In addition, the overall bug submission limit is reduced to 5, as opposed to 10 for the other two problems.
Recall that for \securelog
and \atm, a break constitutes direct evidence that a vulnerability has
been exploited, but not \emph{which} vulnerability, if more than one
is present. As such, if a build-it team were to fix a vulnerability
during the fix-it phase, doing so would not shed light on which breaks
derived from that vulnerability, vs. others. The contest thus limits
break-it teams to one break each for confidentiality and integrity,
per target team. (See \S\ref{sec:problem} and
\S\ref{sec:atm-problem}.)  For \ehr, a security vulnerability is
associated with a test run, so a fix of that vulnerability \emph{will}
unify all breaks that exploit that vulnerability, just as correctness
fixes do. That means we need not impose a limit on them. The
consequences of this design are discussed in \S\ref{sec:analysis}.

\section{Build-it Submissions: Successes and Failures}
\label{sec:stories}
After running a BIBIFI contest, we have all of the code written by
the build-it teams, and the bug reports submitted by the break-it teams.
Looking at these artifacts, we can get a sense of what build-it teams did right,
and what they did wrong. This section presents a sample of success and failure
stories, while \S\ref{sec:analysis} presents a broader statistical
analysis that suggests overall trends.

\subsection{Success Stories}
The success stories bear out some old chestnuts of wisdom in the security community:
submissions that fared well through the break-it phase made heavy use of existing 
high-level cryptographic libraries with few ``knobs'' that allow for
incorrect usage~\cite{bernstein2012security}. Similarly, successful
submissions limited the size and location of security-critical code~\cite{owaspchecklist}.

\paragraph*{\atm}
One implementation of the \atm problem, written in Python, made use of the SSL PKI infrastructure. The
implementation used generated SSL private keys to establish a root of trust that authenticated the 
\texttt{atm} program to the \texttt{bank} program. Both the \texttt{atm} and \texttt{bank} required that the connection be signed with 
the certificate generated at runtime. Both the \texttt{bank} and the \texttt{atm} implemented their communication 
protocol as plain text then wrapped in HTTPS\@. To 
find bugs in this system, other contestants would need to break the security of OpenSSL. 

Another implementation, written in Java, used the NaCl library. 
This library intentionally provides a very high level API to ``box'' and ``unbox'' secret values, freeing
the user from dangerous choices. As above, to break this system, other contestants would need to 
break NaCl, first. 

\paragraph*{\securelog}
An implementation of the \securelog problem, written in Java, achieved success using a high
level API\@. They used the BouncyCastle library to construct a valid encrypt-then-MAC scheme over
the entire log file. 
BouncyCastle allowed them to easily authenticate the whole log file,
protecting them from integrity attacks that swapped the order of
encrypted binary data in the log. 

\paragraph*{\ehr}
The most successful solutions for the \ehr problem were localized access control logic checks to
a single function with a general interface, rather repeating checking
code for each command that needed it. Doing so reduced the likelihood
of a mistake. One of the most successful teams used a fairly complex graphical representation of access control 
rules, but by limiting the number of places this graph was manipulated
they could efficiently and correctly check access rights without introducing vulnerabilities.

\subsection{Failure Stories}
The failure modes for build-it submissions are distributed along a spectrum ranging from 
``failed to provide any security at all'' to ``vulnerable to extremely subtle timing attacks.'' 
This is interesting because a similar dynamic is observed in the software marketplace today. 

\paragraph*{\securelog}
Many implementations of the \securelog problem failed to use
encryption or authentication codes, presumably because the builders
did not appreciate the need for them.
Exploiting these design flaws was trivial for break-it teams. Sometimes log data
was written as plain text, other times log data was serialized using the Java object serialization
protocol. 

One break-it team discovered a privacy flaw which they could exploit with at most fifty probes. 
The target submission truncated the authentication token (i.e., the key) so that it was vulnerable to a brute force attack. 

Some failures were common across \securelog implementations: if an implementation used 
encryption, it might not use authentication. If it used authentication, it would authenticate
records stored in the file individually, not globally. The implementations would also
relate the ordering of entries in the file to the ordering of events in time, allowing for
an integrity attack that changes history by re-ordering entries in the file. 

\paragraph*{\atm}
The \atm problem allows for interactive attacks (not possible for the log), and the attacks became cleverer 
as implementations used cryptographic constructions incorrectly. One implementation used cryptography,
but implemented RC4 from scratch and did not add any randomness to the key or the cipher stream. An 
attacker observed that the ciphertext of messages was distinguishable and largely unchanged from 
transaction to transaction, and was able to flip bits in a message to change
the withdrawn amount. 

Another implementation used encryption with authentication, but did
not use randomness; as such error messages were always distinguishable
from success messages. An attack was constructed against 
this implementation where the attack leaked the bank balance by observing different withdrawal
attempts, distinguishing the successful from failed transactions, and performing a binary search to identify
the bank balance given a series of withdraw attempts. 

Some failures were common across \atm problem implementations. Many implementations kept the key fixed
across the lifetime of the \texttt{bank} and \texttt{atm} programs and did not use a nonce in the messages. 
This allowed attackers to replay messages freely between the
\texttt{bank} and the \texttt{atm}, violating integrity via
unauthorized withdrawals. Several implementations used encryption, but
without authentication. (This sort of mistake has been observed in
real-world ATMs, as has, amazingly, a complete lack of encryption use~\cite{atm-hacks}.)
These implementations used a library such as OpenSSL, the Java cryptographic framework, or the Python
pycrypto library to have access to a symmetric cipher such as AES, but
either did not use these libraries at a level where authentication was provided in addition to encryption,
or they did not enable authentication. 

\paragraph*{\ehr}
The \ehr problem asks participants to consider a complex logical security problem.  In this scenario, the specification was much more complicated. 
Only one team used a library
to abstract away the access control problem.  Instead, most participants developed a system of checks themselves to ensure everything was handled correctly. This led to a variety of failures when participants were unable to cover all possible security edge cases.  

In some instances, vulnerabilities were introduced because they did not properly implement the specification. 
Some participants hardcoded passwords making them easily guessable by an attacker. 
Other participants 
did not include checks for the delegation command to ensure that the principal had the right to delegate along 
with the right they were trying to delegate.

Other participants failed to consider the security implications of their design decisions when the specification did 
not provide explicit instructions. For example, many of the participants did not check the delegation chain back to 
its root.  Therefore, once a principal received an access right, they maintained this right even if it no longer 
belonged to the principal that delegated it to them.

Finally, other teams simply made errors when implementing the access control logic.  In some cases, these 
mistakes introduced faulty logic into the access control checks.  One team made a mistake in their control flow 
logic such that if a principal had no delegated rights, the access control checks were skipped---because a lookup error would have occurred.  In other cases, these mistakes led to uncaught runtime errors that allowed the attacker to crash the server, making it unavailable to other users.

\paragraph*{Discussion}
Interestingly, we did not observe any sophisticated attacks used by the breakers.  The submitted breaks were 
simply edge-case tests.  These issues show that even given a very detailed specification, it can be difficult for 
developers to remember and test for all possible problems. 

As a corpus for research, this data set is of interest for future mining. What common design
patterns were used and how did they impact the outcome? Are there any metrics we can extract
from the code itself that can predict security scores during break-it? We defer this analysis
to future work.

\section{Quantitative Analysis}
\label{sec:analysis}
This section quantitatively analyzes data we gathered from our 2015
and 2016 contests.\footnote{We also ran a contest during Fall 2014~\cite{bibifi-cset15} but
  exclude it from consideration due to differences in how it was
  administered.}
We consider participants' performance in each phase of the
contest, including which factors contribute to high scores after the 
build-it round,
resistance to breaking by other teams, and strong performance as
breakers. 

We find that on average, teams that program using statically-typed languages 
are $11\times$ less likely to have security bugs identified in their code
compared to those using C and C++.
Success in breaking, 
and particularly in identifying security bugs in other teams' code, is 
correlated with having more team members, as well as with
participating successfully in 
the build-it phase (and therefore having given thought to how to 
secure an implementation). Somewhat surprisingly, use of advanced techniques 
like fuzzing and static analysis did not appear to affect breaking success.  
Overall, integrity bugs were far more
common than privacy bugs or crashes. 
%
%
The contests that used the \atm problem and the \ehr problem were
associated with more security bugs than the \securelog contest. 

\subsection{Data collection}

For each team, we collected a variety of observed and self-reported
data. When signing up for the contest, teams reported standard
demographics and features such as coding experience and programming
language familiarity. After the contest, each team member optionally
completed a survey about their performance. In addition, we extracted
information about lines of code written, number of commits, etc. from
teams' Git repositories.

Participant data was anonymized and stored in a manner approved by our
institution's human-subjects review board. 
Participants consented to
have data related to their activities collected, anonymized, stored,
and analyzed. 
A few participants did not consent to research involvement, 
so their personal data was not used in the data analysis. 

\subsection{Analysis approach}
To examine factors that correlated with success in building and
breaking, we apply regression analysis. 
Each regression model attempts to explain some outcome variable using
one or more measured factors. 
For most outcomes, such as participants' scores, we can use ordinary
linear regression, which estimates how many points a given factor
contributes to (or takes away from) a team's score. 
To analyze binary outcomes, such as whether or not a security bug was
found, we apply logistic regression, which estimates how each factor impacts the likelihood of
an outcome. 

We consider many variables that could potentially impact teams' results.
To avoid over-fitting, we select as potential factors those
variables that we believe are of most interest, within acceptable
limits for power and effect size.
As we will detail later, we use the same factors as the analysis
in our earlier conference paper~\cite{Ruef:2016:BBF:2976749.2978382},
plus one more, which identifies participation in the added 
contest (\ehr). The impact of the added data on the analysis, compared
to the analysis in the earlier paper, is considered in
Section~\ref{subsec:modeldifferences}. 
We test models with all possible combinations of our chosen
potential factors and select the model with the minimum Akaike
Information Criterion (AIC)~\cite{burnham2011aic}. 
The final models are presented.

Each model is presented as a table with each factor as well as the \emph{p}-value for that factor. 
Significant \emph{p}-values (< 0.05) are marked with an asterisk. 
Linear models include the coefficient relative to the baseline factor
and the 95\% confidence interval. 
Logistic models also include the exponential coefficient 
and the 95\% confidence interval for the exponential coefficient. 
The exponential coefficient indicates how many times more likely 
the measured result occurs
relative to the baseline factor. 

We describe the results of each model below. This was not a completely
controlled experiment (e.g., we do not use 
random assignment), so our models demonstrate correlation 
rather than causation. Our observed effects may involve confounds, and many 
factors used as independent variables in our data are correlated with each other. 
This analysis also assumes that the factors we examine have linear effect on participants' scores 
(or on likelihood of binary outcomes); while this may not be the case in reality, it is a common 
simplification for considering the effects of many factors. We also note that some of 
the data we analyze is self-reported, so may not be entirely precise (e.g., some participants 
exaggerating about which programming languages they know); however, minor 
deviations, distributed across the population, act like noise and have little impact 
on the regression outcomes. 

\subsection{Contestants}

We consider three contests offered at different times:

\textbf{\securelog}: We held one contest using the
\securelog problem (\S\ref{sec:problem}) during May--June 2015
  as the capstone to a Cybersecurity MOOC sequence.\footnote{\url{https://www.coursera.org/specializations/cyber-security}} 
  Before completing in the capstone, participants passed courses on
  software security, cryptography, usable security, and hardware
  security. 

\textbf{\atm}: During Oct.--Nov. 2015 we offered the \atm problem  (\S\ref{sec:atm-problem}) as two contests
  simultaneously, one as a MOOC capstone, and the other open to
  U.S.-based graduate and undergraduate
   students. We merged the contests after the build-it phase,
  due to low participation in the open contest.
MOOC and open participants
were ranked independently to determine grades and prizes. 

\textbf{\ehr}: In Sep.--Oct. 2016 we ran one constest offering
  the \ehr problem (\S\ref{sec:ehr}) open to both MOOC capstone
  participants as well as graduate and undergraduate students.  

The U.S. was the most represented country in our contestant pool, but
was not the majority. 
There was also representation from developed countries with a reputation
both for high technology and hacking acumen. Details of the most 
popular countries of origin can be found in Table~\ref{tab:demo-country}, 
and additional information about contestant demographics is presented 
in Table~\ref{tab:demog}. In total, 156 teams participated in either
the build-it or break-it phases, most of which participated in both.


\begin{table}[t]
\centering
\small
\begin{tabular}{lrrrrr}
\toprule
{\bf Contest}     & \textbf{USA} & \textbf{India} & \textbf{Russia} & \textbf{Brazil} & \textbf{Other} \\
\midrule
Spring 2015 & 30  & 7     & 12     & 12     & 120   \\
Fall 2015   & 64  & 14    & 12     & 20     & 110  \\
Fall 2016   & 44 & 13 & 4 & 12 & 103 \\
\bottomrule
\end{tabular}
\caption{Contestants, by self-reported country.}
\label{tab:demo-country}
\vspace{-3ex}
\end{table}

\begin{table}[t]
\centering
\small
\label{my-label}
\begin{tabular}{p{.3\columnwidth}|llll}
  \toprule
\textbf{Contest} & \textbf{Spring 15} & \textbf{Fall 15} & \textbf{Fall 16} \\ 
\midrule
Problem   & \securelog & \atm & \ehr \\
  \# Contestants                                       & 156            & 145 & 105                    \\
\% Male                                            & 91 \% & 91 \% & 84 \%               \\
\% Female					&  5 \% & 8 \% & 4 \% \\
Age {\tiny (mean/min/max)}                       & 34.8/20/61     & 32.2/17/69 & 29.9/18/55         \\
				\% with CS degrees                                   & 35 \%             & 35 \% & 39 \%                  \\ 
				Years programming                                 & 9.6/0/30       & 9.4/0/37 & 9.0/0/36              \\ 
\# Build-it teams               & 61             & 40 & 29                    \\
Build-it team size                                & 2.2/1/5        & 3.1/1/6 & 2.5/1/8              \\
\# Break-it teams (that also built)              & 65 (58)           & 43 (35) & 33 (22)                   \\
Break-it team size                                & 2.4/1/5        & 3.1/1/6 & 2.6/1/8            \\ 
\# PLs known per team                & 6.8/1/22       & 9.1/1/20      & 7.8/1/17         \\
\% MOOC & 100 \% & 84 \% & 65 \% \\ 
\bottomrule
\end{tabular}
\caption{Demographics of contestants from qualifying
	teams. Some participants 
  declined to specify gender.}
\label{tab:demog}
\vspace{-4ex}
\end{table}


\begin{figure}[t!]
\centering
  \includegraphics[width=0.75\columnwidth]{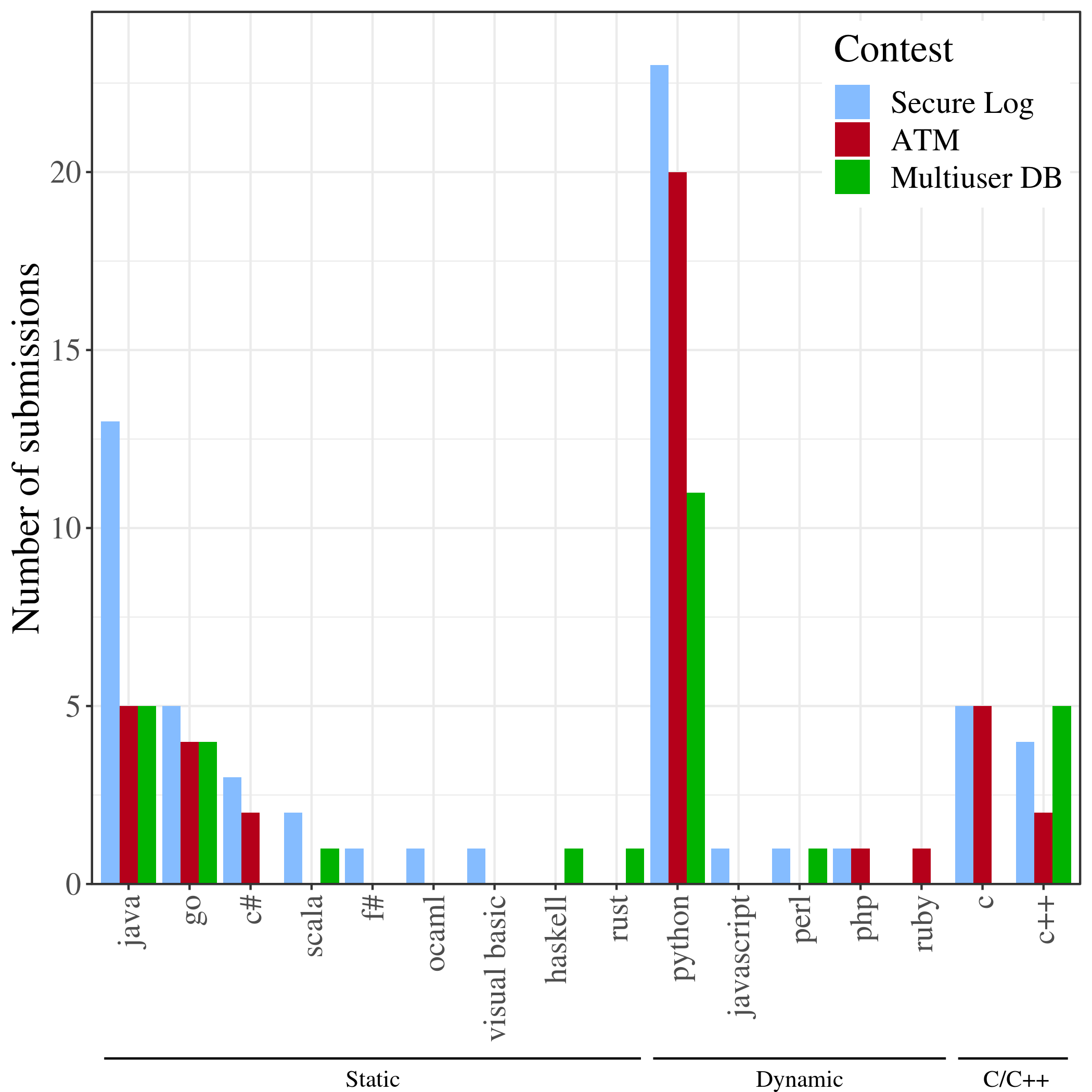}
\caption[]{The number of build-it submissions in each contest, organized by primary programming 
				language used. The languages are grouped by category.} 
\label{fig:submissions-by-language}
\vspace{-1ex}
\end{figure}

\subsection{Ship scores}
\label{ss:ship-it}

We first consider factors correlating with a team's \emph{ship} score,
which assesses their submission's quality
before it is attacked by the other teams (\S\ref{sec:design}).  This data set contains all
130 teams from the \securelog, \atm, and \ehr contests that qualified
after the build-it phase. The contests have nearly the same number of
correctness and performance tests, but different numbers of
participants. We set the constant multiplier $M$ to be 50
for the contests, which effectively normalizes the scores.

\paragraph*{Model setup}

To ensure enough statistical power to find meaningful relationships, our modeling 
was designed for a prospective effect size roughly equivalent to 
Cohen's {\it medium} effect heuristic, $f^2=0.15$~\cite{Cohen:1988}.
An effect of this size corresponds to a coefficient of determination $R^2=0.13$, 
suggesting we could find an effect if our model can explain at least 13\% of 
the variance in the outcome variable. We report the observed coefficient of 
determination for the final model with the regression results below.

As mentioned above, we reuse the factors chosen for the analysis in our
earlier paper~\cite{Ruef:2016:BBF:2976749.2978382}. Their number was
guided by a power analysis of the 
contest data we had at the time, which involved the $N=101$
build-it teams that participated in \securelog and \atm.
With an assumed power of $0.75$, the power analysis suggested we
  limit the covariate factors used in our model to nine
degrees of freedom, which yields a prospective $f^2=0.154$. 
With the addition of \ehr data, we add one more factor, which is
choice of \ehr as an option for which contest the submission 
belongs to. This adds a 10th degree of freedom, as well as 29 additional teams for 
a total $N=130$. At $0.75$ power, this yields a prospective $f^2=0.122$, which 
is better than in the earlier paper's analysis. 

We selected the factors listed in Table \ref{tab:buildfactors}. 
\emph{Knowledge of C} is included as a proxy
for comfort with low-level implementation details, a skill often
viewed as a prerequisite for successful secure building or breaking.
\emph{\# Languages known} is how many unique programming languages
team members collectively claim to know (see the second to last row of
Table~\ref{tab:demog}). For example, on a two-member 
team where member A claims to know C++, Java, and Perl and member B
claims to know Java, Perl, Python, and Ruby, the language count would
be 5. 
\emph{Language category} is the ``primary" language category we manually identified in
each team's submission.
Languages were categorized as either C/C++, statically-typed (e.g., Java, Go,
but not C/C++), or dy\-nam\-ically-typed (e.g., Perl, Python). 
Precise category allocations, and total
submissions for each language, segregated by contest, are given in
Figure~\ref{fig:submissions-by-language}.

\begin{table}
\begin{center}
\small
\begin{tabular}{l p{8cm} l}
\toprule
\textbf{Factor} & \textbf{Description} & \textbf{Baseline} \\
\midrule
Contest & \securelog, \atm, or \ehr contest. & \securelog \\
\# Team members & A team's size. & --- \\
Knowledge of C & The fraction of team members who know C or C++. & --- \\
\# Languages known & Number of programming languages team members know. & --- \\
Coding experience & Average years of programming experience. & --- \\
Language category & C/C++, statically-typed, or dy\-nam\-ically-typed language. & C/C++ \\
Lines of code & Lines of code count for the team's final submission. & --- \\
MOOC & If the team was participating in the MOOC capstone. & non-MOOC \\
\bottomrule
\end{tabular}
\caption{Factors and baselines for build-it models.}
\label{tab:buildfactors}
\end{center}
\vspace{-4ex}
\end{table}

\paragraph*{Results}

\begin{figure}[t!]
\begin{center}
  \includegraphics[width=0.85\columnwidth]{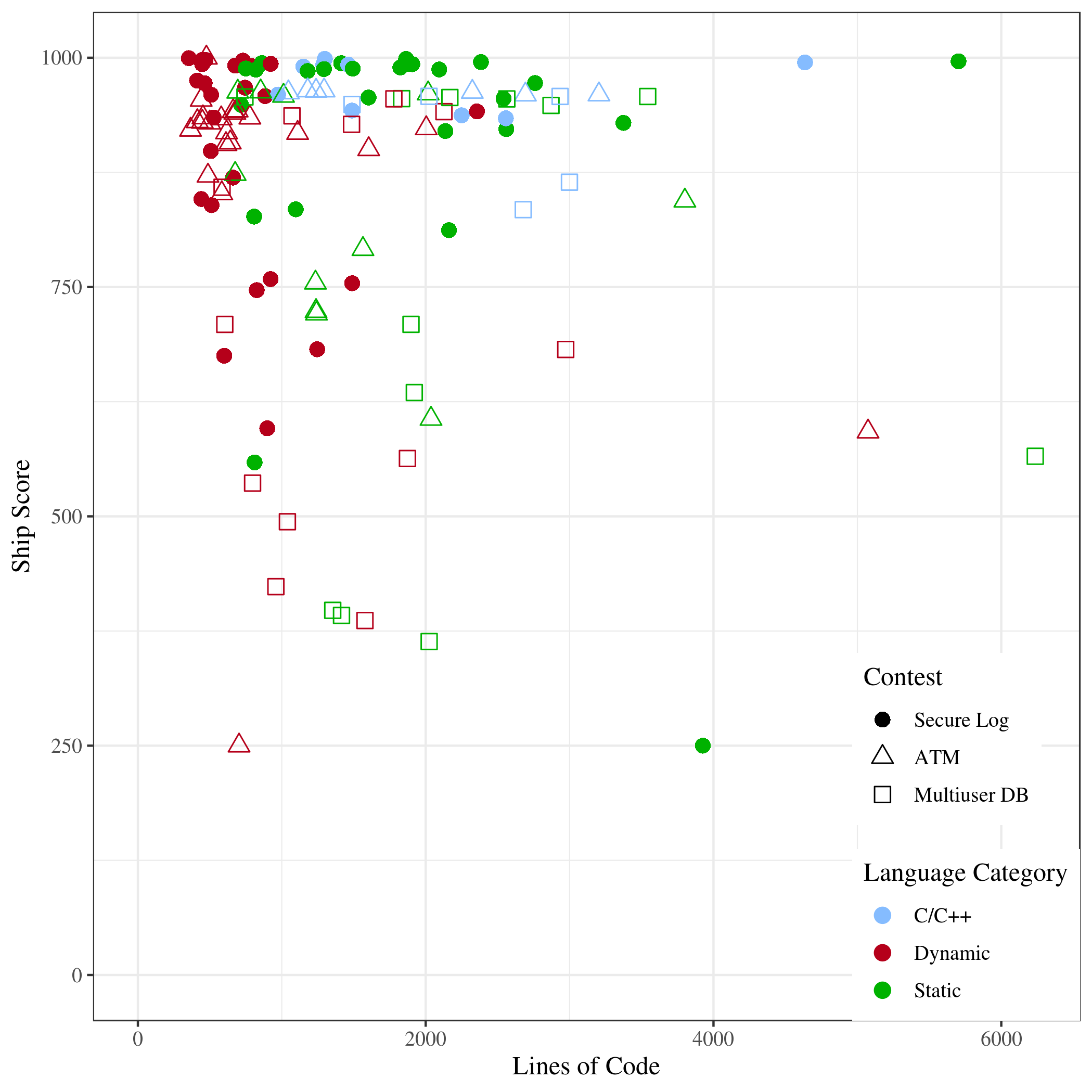}
\end{center}
\caption[]{Each team's ship score, compared to the lines of code in its implementation and 
organized by language category and contest. 
Using C/C++ correlates with a higher ship score.}
\label{fig:LOC-langCat-ship}
\end{figure}

The final model (Table~\ref{tab:ship-model}) with $R^2 = 0.232$ 
captures almost $\frac{1}{4}$ of the variance. We find this number encouraging
given how relatively uncontrolled the environment is and how many 
contributing, but unmeasured, factors there could be.
Our regression results indicate that
ship score is strongly correlated with
language choice. Teams that programmed in C or C++ performed on
average 133 and 112 points better than those who programmed in
dynamically typed or statically typed languages,
respectively. Figure~\ref{fig:LOC-langCat-ship} illustrates that while
teams in many language categories performed well in this phase, only
teams that did not use C or C++ scored poorly. 

The high scores for C/C++ teams could be due to better scores on
performance tests and/or due to implementing optional features. We
confirmed the main cause is the former. 
Every C/C++ team for the \securelog contest implemented all optional
features, while six teams in the other categories implemented only six
of ten and one team implemented none; the \atm contest offered no
optional features; 
for the \ehr contest, four C/C++ teams implemented all optional features while one C/C++ team implemented five of nine. 
We artificially increased the scores of all
teams as if they had implemented all optional features and reran the
regression model. 
In the resulting model, 
the difference in coefficients between C/C++ and the other language categories dropped only slightly. 
This indicates that the majority of improvement in C/C++ ship score comes from performance. 


The number of languages known by a team is not quite statistically significant, but the confidence interval in the model suggests that each programming language known increases ship scores by between 0 and 12 points. 
Intuitively, this makes sense since contestants that know more languages have more programming experience and have been exposed to different paradigms. 

Lines of code is also not statistically significant, but the model hints that each additional line of code in a team's
submission is associated with a minor drop in ship score. 
Based on our qualitative observations (see~\S\ref{sec:stories}), we
hypothesize this may relate to more reuse of code from libraries,
which frequently are not counted in a team's LOC (most libraries were installed directly
on the VM, not in the submission itself).  
We also found that, as further noted
above, submissions that used libraries with more sophisticated,
lower-level interfaces tended to have more code and more
mistakes; i.e., more steps took place in the application (more code) but
some steps were missed or carried out incorrectly (less secure/correct).
Figure~\ref{fig:LOC-langCat-ship} shows that LOC is (as
expected) associated with the category of language being used. While
LOC varied widely within each language type, dynamic submissions were
generally shortest, followed by static submissions and then those
written in C/C++ (which has the largest minimum size).\footnote{%
Our earlier model for the \securelog and \atm contests 
found that lines of code was actually statistically significant. 
We discuss this further in \S\ref{subsec:modeldifferences}. }

\begin{table}[t]
\begin{center}
\small
\begin{tabular}{l r r r }
\toprule
\textbf{Factor} & \textbf{Coef.} & \textbf{CI} & \textbf{$p$-value} \\
\midrule
\securelog & --- & --- & ---\ahs{}\\
\color{insignificant} \atm  &  \color{insignificant} -47.708  &  \color{insignificant} [-110.34, 14.92]  &  \color{insignificant} 0.138\ahs{} \\
\significant{\ehr}  &  \significant{-163.901}  &  \significant{[-234.2, -93.6]}  &  \significant{<0.001*} \\
\midrule
C/C++ & --- & --- & ---\ahs{}\\
\significant{Statically typed}  &  \significant{-112.912}  &  \significant{[-192.07, -33.75]}  &  \significant{0.006*} \\
\significant{Dynamically typed}  &  \significant{-133.057}  &  \significant{[-215.26, -50.86]}  &  \significant{0.002*} \\
\midrule
\color{insignificant} \# Languages known  &  \color{insignificant} 6.272  &  \color{insignificant} [-0.06, 12.6]  &  \color{insignificant} 0.054\ahs{} \\
\midrule
\color{insignificant} Lines of code  &  \color{insignificant} -0.023  &  \color{insignificant} [-0.05, 0.01]  &  \color{insignificant} 0.118\ahs{} \\
\bottomrule
\end{tabular}
\end{center}
\caption[]{Final linear regression model of teams' ship scores, indicating how many points 
each selected factor adds to the total score. 
$R^2 = 0.232$.}

\label{tab:ship-model}
\vspace{-2ex}
\end{table}


\subsection{Resilience}
\label{sec:resilience}

Now we turn to measures of a build-it submission's quality, starting
with \emph{resilience}. Resilience is a non-positive score
that derives from break-it teams' bug reports, which are accompanied
by test cases that prove the presence of
defects. The overall build-it score is the
sum of ship score, just discussed, and resilience.
Builders may increase the resilience component during the fix-it phase, as
fixes prevent double-counting bug reports that identify the same defect
(see~\S\ref{sec:design}). 

\begin{table}[t]
\begin{center}
\small
\begin{tabular}{l r r r }
\toprule
				& \textbf{\securelog} & \textbf{\atm} & \textbf{\ehr} \\
\midrule
				Bug reports submitted & 24,796 & 3,701 & 3,749\\
				Bug reports accepted & 9,482 & 2,482 & 2,046 \\
				Fixes submitted & 375 & 166 & 320 \\
				Bug reports addressed by fixes & 2,252 & 966 & 926 \\
\bottomrule
\end{tabular}
\end{center}
\caption{Break-it teams in each contest submitted bug reports, which were judged by the 
automated oracle. Build-it teams then submitted fixes, each of which could potentially 
address multiple bug reports. 
}
\label{tab:bugs-fixes}
\vspace{-2ex}
\end{table}

\begin{figure}[t]
\begin{centering}
\includegraphics[width=\columnwidth]{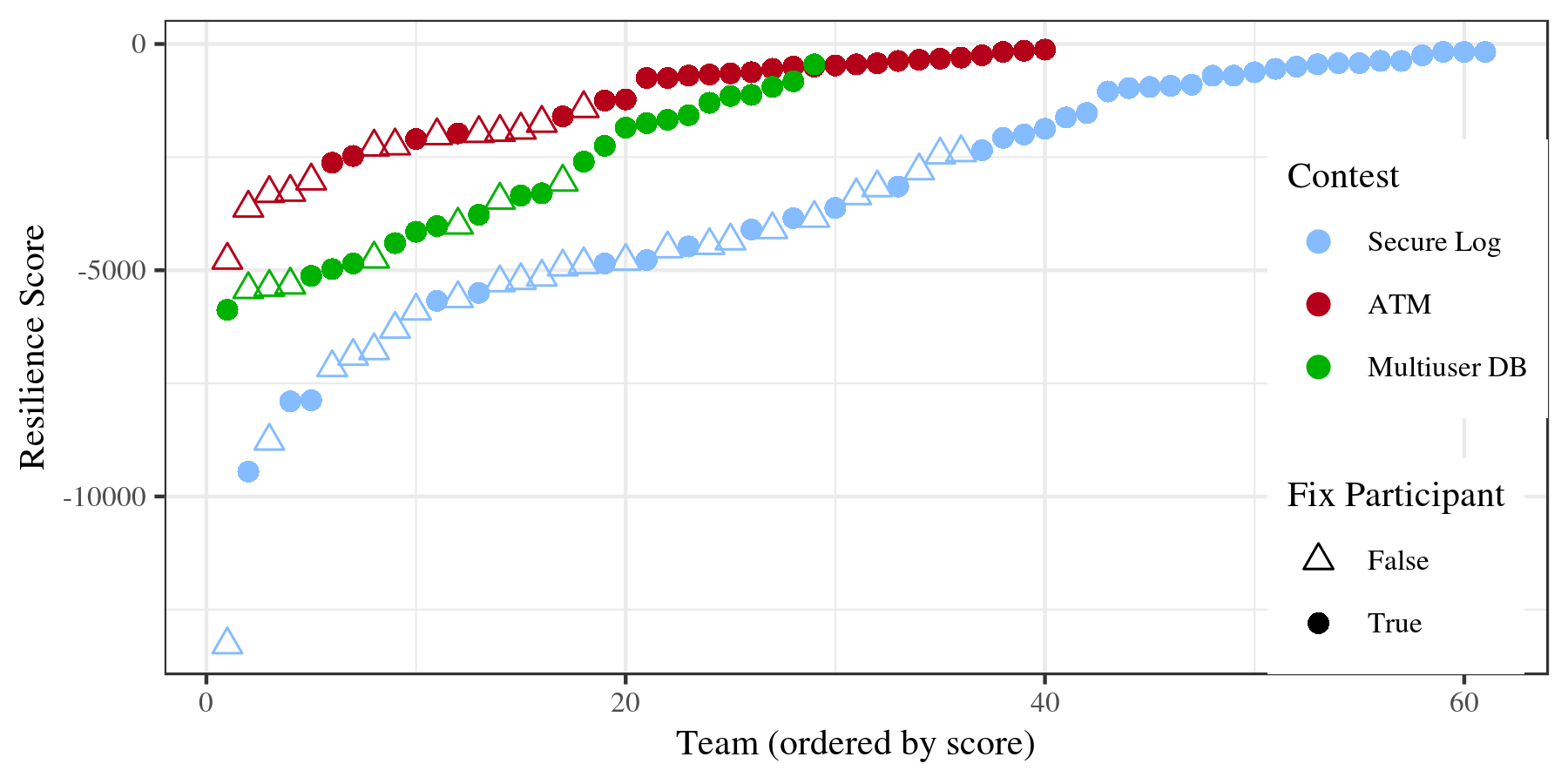}
\caption{Final resilience scores, ordered by team, and plotted for
  each contest problem. Build-it teams who did not bother to fix bugs
  generally had lower scores.}
\label{fig:p3}
\end{centering}
\end{figure}

Unfortunately, upon studying the data we found that a large percentage
of build-it teams opted not to fix any bugs reported against their code,
forgoing the scoring advantage of doing so. We can see this in
Figure~\ref{fig:p3}, which graphs the resilience scores (Y-axis) of
all teams, ordered by score, for the three contests. The circles in the
plot indicate teams that fixed at least one bug, whereas the triangles 
indicate teams that fixed no bugs. We can see that, overwhelmingly,
the teams with the lower resilience scores did not fix any bugs. 
Table~\ref{tab:bugs-fixes} digs a little further into the
situation. It shows that of the bug reports deemed acceptable by the
oracle (the second row), submitted fixes (row 3) addressed only 23\%
of those from the \securelog contest, 38\% of those from the \atm contest, and 45\% of those from the \ehr contest (row 4
divided by row 2). 
It turns out that when counting only ``active" fixers who fixed at least one bug, these averages 
were 56.9\%, 72.5\%, and 64.6\% respectively. 



\paragraph*{Incentivizing fixing.}
This situation is disappointing, as we cannot treat resilience score as
a good measure of code quality (when added to ship score). 
After the first two contests, 
we hypothesized that participants were not sufficiently incentivized to
fix bugs, for two reasons. First, teams that were sufficiently far from
the lead may have chosen to fix no bugs because winning was
unlikely. Second, for MOOC students, once a minimum score is
achieved they were assured to pass; it may be that fixing (many) bugs was
unnecessary for attaining this minimum score. 

In an attempt to more strongly incentivize all teams to fix all (duplicated) bugs, 
we modified the prize structure for the \ehr contest. 
Instead of only giving away prizes to top teams, 
non-MOOC participants could still win monetary prizes if they scored outside of third place. 
Placements were split into different brackets and one team from each bracket was randomly selected to receive the prize. 
Prizes increased based on bracket position 
(\textit{ex,} the fourth and fifth place bracket winner received \$500, while the sixth and seventh place bracket winner received \$375). 
%
%
%
Our hope was that 
builders would submit fixes to bump themselves into a higher bracket which would have a larger payout. 
Unfortunately, it does not appear that fix participation increased for non-MOOC participants for the \ehr contest. 
To confirm this, 
we ran a linear regression model, but according to the model, incentive structure was not a factor in fix participation. 
The model did confirm that teams with a higher score at the end of break-it fixed a greater percentage of the bugs against them. 


\subsection{Presence of security bugs}
\label{sec:security-bug}

While resilience score is not sufficiently meaningful, a useful alternative is
the likelihood that a build-it submission contains a security-relevant
bug; by this we mean any submission against which at least one 
crash, privacy, integrity, or availability defect is demonstrated. In this model we used logistic regression
over the same set of factors as the ship model. 

Table~\ref{tab:secBug-model} lists the results of this logistic regression; 
the coefficients represent the change in log likelihood associated with each 
factor. Negative coefficients indicate lower likelihood of finding a security bug. 
For categorical factors, the exponential of the coefficient (\emph{exp(coef)}) indicates 
how strongly that factor being true affects the likelihood relative to the baseline 
category.\footnote{In cases (such as the \atm contest) where the rate of 
security bug discovery is close to 100\%, the change in log likelihood starts to approach 
infinity, somewhat distorting this coefficient upwards.} For numeric factors, the 
exponential indicates how the likelihood changes with each unit change in that factor. 
%
$R^2$ as traditionally understood does not make sense for a logistic regression. 
There are multiple approximations proposed in the literature, each of which have various pros and cons. 
We present 
Nagelkerke ($R^2 = 0.619$) which suggests the model explains an estimated 61\% of variance \cite{10.1093/biomet/78.3.691}. 

\originalmodel{
\begin{table}[t]
\begin{center}
\small
\begin{tabular}{l r r r r}
\toprule
\textbf{Factor} & \textbf{Coef.} & \textbf{Exp(coef)} & \textbf{CI} & \textbf{$p$-value} \\
\midrule
Fall 2015  &  5.692  &  296.395  &  [20.05, 4381.21]  &  <0.001* \\
\# Languages known  &  -0.184  &  0.832  &  [0.7, 0.98]  &  0.032* \\
Lines of code  &  0.001  &  1.001  &  [1, 1]  &  0.03* \\
\color{Gray} Dynamically typed  &  \color{Gray} -0.751  &  \color{Gray} 0.472  &  \color{Gray} [0.08, 2.64]  &  \color{Gray} 0.393~ \\
Statically typed  &  -2.138  &  0.118  &  [0.02, 0.67]  &  0.016* \\
\color{Gray} MOOC  &  \color{Gray} 2.872  &  \color{Gray} 17.674  &  \color{Gray} [0.67, 467.93]  &  \color{Gray} 0.086~ \\
\bottomrule
\end{tabular}
\end{center}
\caption{ORIGINAL Final logistic regression model, measuring log likelihood of a security bug 
being found in a team's submission. 
}
\label{tab:secBug-orig-model}
\vspace{-2ex}
\end{table}
}

\begin{table}[t]
\begin{center}
\small
\begin{tabular}{l r r r r}
\toprule
\textbf{Factor} & \textbf{Coef.} & \textbf{Exp(coef)} & \textbf{Exp CI} & \textbf{$p$-value} \\
\midrule
\securelog & --- & --- & --- & ---\ahs{}\\
\significant{\atm}  &  \significant{4.639}  &  \significant{103.415}  &  \significant{[18, 594.11]}  &  \significant{<0.001*} \\
\significant{\ehr}  &  \significant{3.462}  &  \significant{31.892}  &  \significant{[7.06, 144.07]}  &  \significant{<0.001*} \\
\midrule
C/C++ & --- & --- & --- & ---\ahs{}\\
\significant{Statically typed}  &  \significant{-2.422}  &  \significant{0.089}  &  \significant{[0.02, 0.51]}  &  \significant{0.006*} \\
\color{insignificant} Dynamically typed  &  \color{insignificant} -0.99  &  \color{insignificant} 0.372  &  \color{insignificant} [0.07, 2.12]  &  \color{insignificant} 0.266\ahs{} \\
\midrule
\color{insignificant} \# Team members  &  \color{insignificant} -0.35  &  \color{insignificant} 0.705  &  \color{insignificant} [0.5, 1]  &  \color{insignificant} 0.051\ahs{} \\
\midrule
\color{insignificant} Knowledge of C  &  \color{insignificant} -1.44  &  \color{insignificant} 0.237  &  \color{insignificant} [0.05, 1.09]  &  \color{insignificant} 0.064\ahs{} \\
\midrule
\color{insignificant} Lines of code  &  \color{insignificant} 0.001  &  \color{insignificant} 1.001  &  \color{insignificant} [1, 1]  &  \color{insignificant} 0.090\ahs{} \\
%
%
\bottomrule
\end{tabular}
\end{center}
\caption{Final logistic model measuring log-likelihood of
  the discovery of a security bug 
in a team's submission. 
Nagelkerke $R^2 = 0.619$.
}
\label{tab:secBug-model}
\vspace{-4ex}
\end{table}

\atm implementations were far more likely than \securelog
implementations to have a discovered security bug.\footnote{This 
coefficient (corresponding to $103\times$) is somewhat exaggerated (see prior footnote), but the 
difference between contests is large and significant.}  
We hypothesize this is due to the increased security design space in the
\atm problem as compared to the \securelog problem. Although it is easier
to demonstrate a security error in the \securelog problem, the \atm
problem allows for a much more powerful adversary (the MITM) that can
interact with the implementation; breakers often took advantage of
this capability, as discussed in~\S\ref{sec:stories}.

\ehr implementations were $31\times$ as likely as \securelog implementations to have a discovered security bug. 
We hypothesize this is due to 
increased difficulty in implementing a custom access control system. 
There are limited libraries available that directly provide the required functionality, 
so contestants needed to implement access control manually, leaving more room for error. 
For the \securelog problem, builders could utilize cryptographic libraries to secure their applications. 
In addition, it was
potentially easier for breakers to discover attacks with the
\ehr problem since they could 
reuse break tests against multiple build submissions.\footnote{%
  One caveat here is that a quirk of the problem definition permitted
  breakers to escalate correctness bugs into security problems by
  causing the state of \ehr submissions to become out of
  sync with the oracle implementation, and behave in a way that seemed
  to violate availability. We only realized after the
  contest was over that these should have been classified as
  correctness bugs. For the data analysis, we retroactively
  reclassified these bugs as correctness problems. Had they been
  classified properly during the contest, break-it team behavior might
  have changed, i.e., to spend more time hunting proper security
  defects. 
}
  
%
%

\begin{figure}[t]
\begin{centering}
\includegraphics[width=.85\columnwidth]{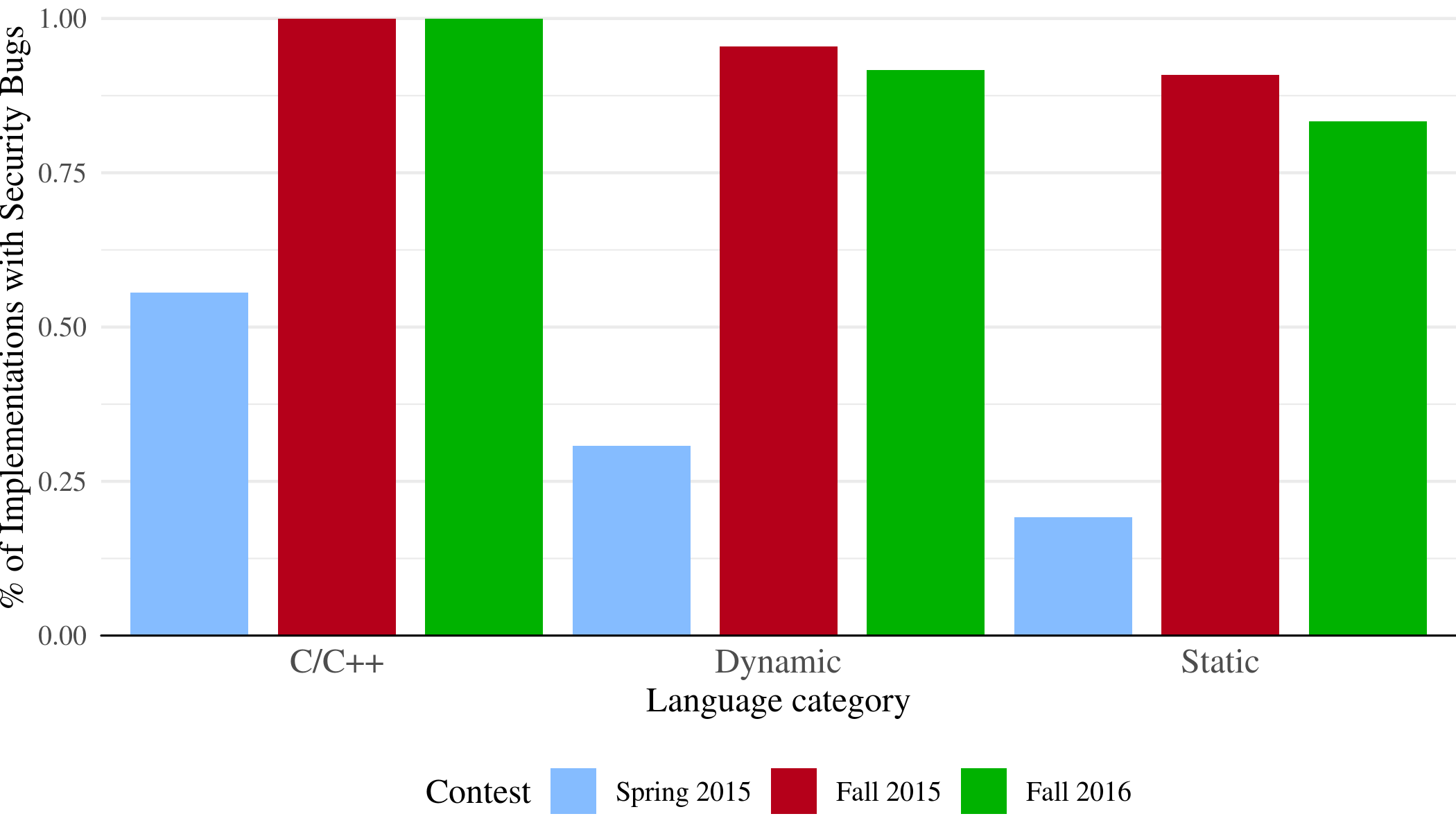} 
\caption[]{The fraction of teams in whose submission a security bug was found, by
contest and language category.}
\label{fig:bugsfound-by-language}
\end{centering}
\end{figure}

The model also shows that C/C++ implementations were more likely to
contain an identified security bug than either static or dynamic
implementations. For static languages, this effect is significant and
indicates that a C/C++ program was about $11\times$ (that is,
$1/0.089$) as likely to contain an identified bug. This effect is
clear in Figure~\ref{fig:bugsfound-by-language}, which plots the
fraction of implementations that contain a security bug, broken down
by language type and contest problem. Of the 21 C/C++ submissions (see
Figure~\ref{fig:submissions-by-language}), 17 of them had a security bug: 5/9 for the \securelog contest,
7/7 for the \atm contest, and 5/5 for the \ehr contest. All five of the buggy implementations from the \securelog contest
had a crash defect, and this was the only
security-related problem for three of them; none of the \atm implementations had
crash defects; four of the \ehr C/C++ submissions had crash defects. 


The model shows four factors that played a role in the
outcome, but not in a statistically significant way: 
using a dynamically typed language, 
lines of code of an implementation, 
developer knowledge of C,
and number of team members. 
We see the
effect of the first in Figure~\ref{fig:bugsfound-by-language}. 
We note that the number of team members is just outside the threshold of being significant. 
This suggests that an implementation is
$1.4\times$ ($1/0.705$) less likely to have a security bug present for each team member.




\subsection{Breaking success}
\label{ss:breaking-success}

Now we turn our attention to break-it team performance, i.e., how
effective teams were at finding defects in build-it teams' submissions.
First, we consider how and why teams performed as indicated by their
(normalized) break-it score \emph{prior to the fix-it phase}. 
We do this to measure a team's raw output, ignoring whether other
teams found the same bug (which we cannot assess with confidence
due to the lack of fix-it phase participation
per~\S\ref{sec:resilience}). 
This data set includes 141 teams that participated in the break-it
phase for the \securelog, \atm, and \ehr contests. 
We also model which factors contributed to \emph{security bug
  count}, or how many total security bugs a break-it team found. 
Doing this disregards a break-it team's effort at finding correctness
bugs.


We model both break-it score and security bug count using several of the same 
potential factors as discussed previously, 
but applied to the breaking team rather than the building team. 
In particular, we include the \emph{Contest} they participated in, whether they were
\emph{MOOC} participants, the number of break-it \emph{Team members}, 
average team-member \emph{Coding experience},
average team-member \emph{Knowledge of C}, and unique
\emph{Languages known} by the break-it team members. We also add two
new potential factors. 1) Whether the breaking team also qualified as a \emph{Build participant}. 2) Whether the breaking team reported using \emph{Advanced techniques} like software analysis
or fuzzing to aid in bug finding. Teams that only used manual inspection and 
testing are not categorized as advanced. 34 break-it teams (24\%) reported using advanced techniques.
These factors are summarized in Table \ref{tab:breakfactors}.

\begin{table}
\begin{center}
\small
\begin{tabular}{l p{8cm} l}
\toprule
\textbf{Factor} & \textbf{Description} & \textbf{Baseline} \\
\midrule
Contest & \securelog, \atm, or \ehr contest. & \securelog \\
\# Team members & A team's size. & --- \\
Knowledge of C & The fraction of team members who know C or C++. & --- \\
\# Languages known & Number of programming languages team members know. & --- \\
Coding experience & Average years of programming experience. & --- \\
MOOC & If the team was participating in the MOOC capstone. & non-MOOC \\
Build participant & If the breaking team qualified as a build participant. & non-builder \\
Advanced techniques & If the breaking team used software analysis or fuzzing. & Not advanced \\
\bottomrule
\end{tabular}
\caption{Factors and baselines for break-it models.}
\label{tab:breakfactors}
\end{center}
\vspace{-3ex}
\end{table}



When carrying out the power analysis for these two
models, we aimed for a {\it medium}
effect size by Cohen's heuristic~\cite{Cohen:1988}.
Assuming a power of 0.75, our conference paper considered a population
of $N=108$ for the \securelog and \atm contests; with
the eight degrees of freedom it yields a prospective effect size $f^2=0.136$. 
Including the \ehr contest increases the degrees of freedom to nine and raises the population to $N=141$. 
This yields a prospective effect size $f^2=0.107$,
which (again) is an improvement over the initial analysis.

\begin{table}[t]
\begin{center}
\small
\begin{tabular}{l r r r }
\toprule
\textbf{Factor} & \textbf{Coef.} & \textbf{CI} & \textbf{$p$-value} \\
\midrule
\securelog & --- & --- & ---\ahs{}\\
\significant{\atm}  &  \significant{-2401.047}  &  \significant{[-3781.59, -1020.5]}  &  \significant{<0.001*} \\
\color{insignificant} \ehr  &  \color{insignificant} -61.25  &  \color{insignificant} [-1581.61, 1459.11]  &  \color{insignificant} 0.937\ahs{} \\
\midrule
\significant{\# Team members}  &  \significant{386.975}  &  \significant{[45.48, 728.47]}  &  \significant{0.028*} \\
\midrule
\color{insignificant} Coding experience  &  \color{insignificant} 87.591  &  \color{insignificant} [-1.73, 176.91]  &  \color{insignificant} 0.057\ahs{} \\
\midrule
\color{insignificant} Build participant  &  \color{insignificant} 1260.199  &  \color{insignificant} [-315.62, 2836.02]  &  \color{insignificant} 0.119\ahs{} \\
\midrule
\color{insignificant} Knowledge of C  &  \color{insignificant} -1358.488  &  \color{insignificant} [-3151.99, 435.02]  &  \color{insignificant} 0.14\ahs{} \\
%
%
\bottomrule
\end{tabular}
\end{center}
\caption{
Final linear regression model of teams' break-it scores, indicating how many points 
each selected factor adds to the total score. 
$R^2 = 0.15$.
}
\label{tab:breakit-model}
\vspace{-4ex}
\end{table}

\paragraph*{Break score} The model considering break-it score is given
in Table~\ref{tab:breakit-model}. 
%
It has a coefficient of determination $R^2 = 0.15$ which is adequate. 
%
The model shows that teams with more
members performed better, with an average of 387 additional points per
team member. Auditing code for errors is
an easily parallelized task, so teams with more members could divide
their effort and achieve better coverage. 

\begin{figure} 
  \begin{subfigure}{\columnwidth}
    \centering
    \includegraphics[width=0.8\columnwidth]{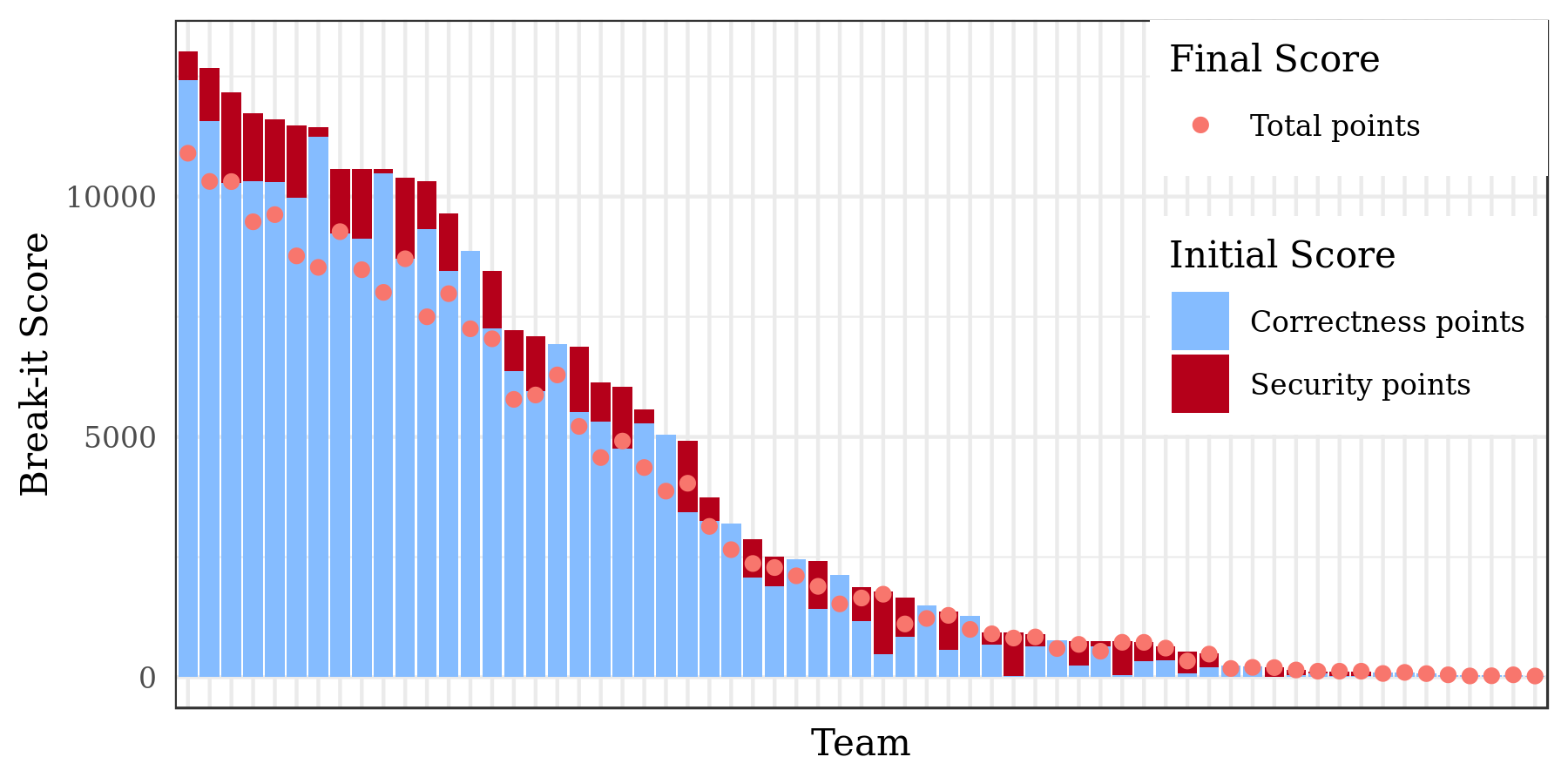}
	\caption{\securelog}
  \end{subfigure}
  \hfill
  \begin{subfigure}{\columnwidth}
    \centering
    \includegraphics[width=0.8\columnwidth]{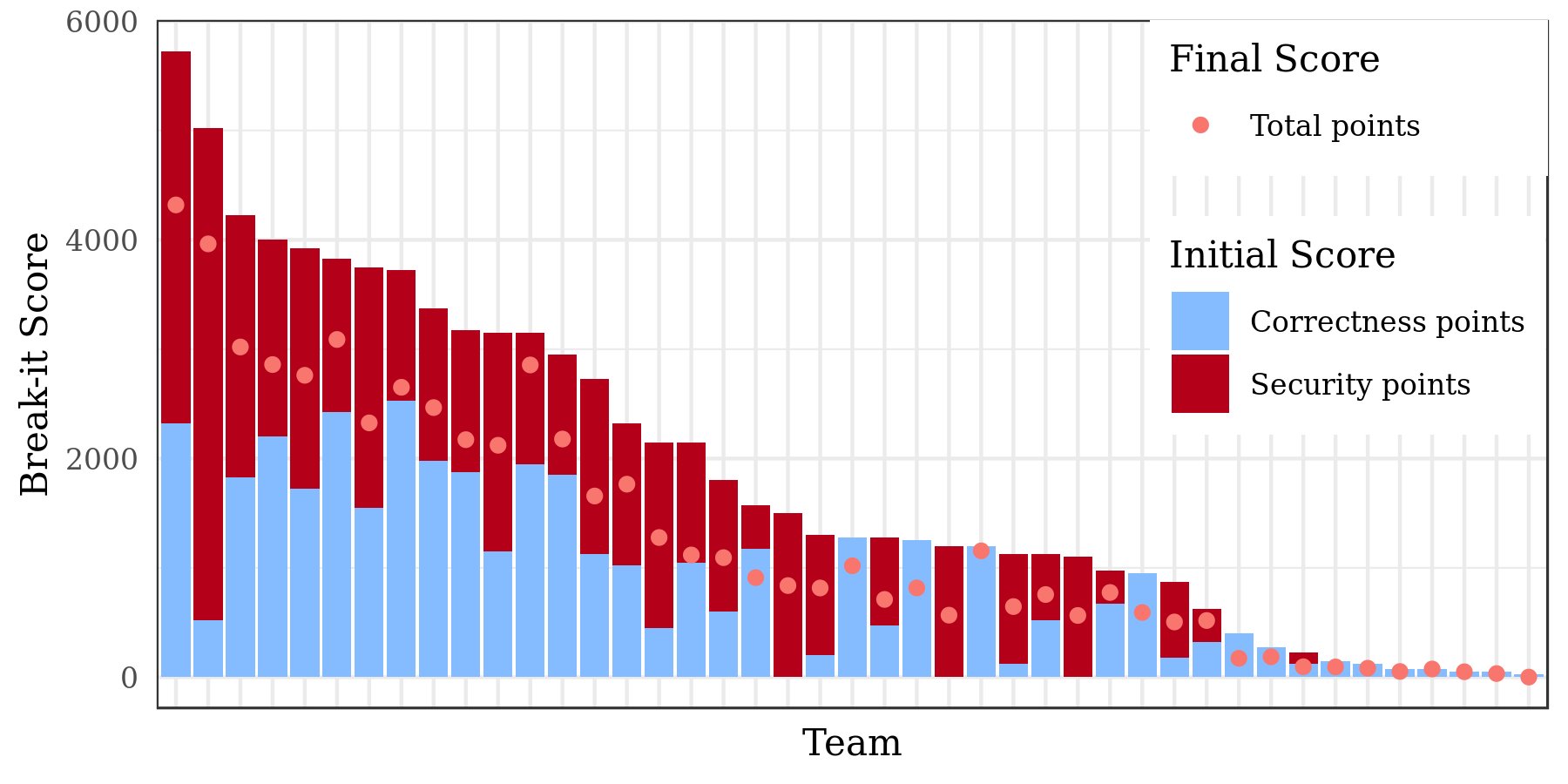}
	\caption{\atm}
  \end{subfigure}
  \hfill
  \begin{subfigure}{\columnwidth}
    \centering
    \includegraphics[width=0.8\columnwidth]{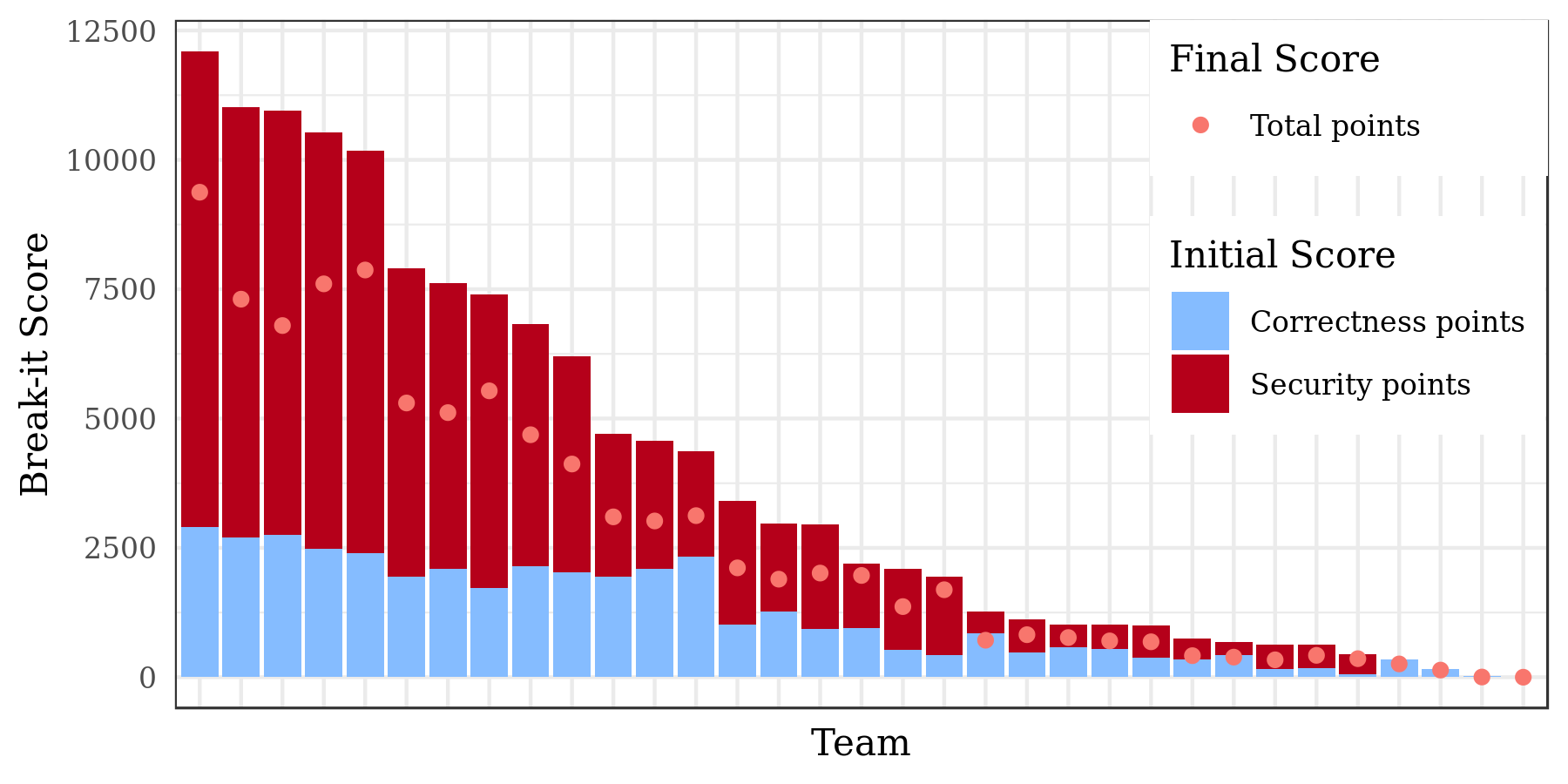}
	\caption{\ehr}
  \end{subfigure}
\caption{\label{fig:security-v-correctness}
Scores of break-it teams prior to the fix-it phase, broken down by points from security 
and correctness bugs. The final score of the break-it team (after fix-it phase) is noted as a dot.
Note the different ranges in the $y$-axes. 
In
general, the \securelog contest had the least proportion of points coming from security breaks. 
}
\end{figure}

The model also indicates that \securelog teams performed
significantly better than \atm teams, and \ehr teams performed similarly to \atm teams.
Figure~\ref{fig:security-v-correctness} illustrates that correctness
bugs, despite being worth fewer points than security bugs, dominate overall 
break-it scores for the \securelog contest. In the \atm contest, the scores 
are more evenly distributed between correctness and security bugs.
This outcome is not surprising to us, as it was somewhat by design.
The \securelog problem defines a rich command-line interface 
with many opportunities for subtle correctness errors that break-it teams can
target. It also allowed a break-it team to submit up to 10 correctness
bugs per build-it submission. To nudge teams toward finding more
security-relevant bugs, we reduced the submission limit from 10 to 5,
and designed the \atm and \ehr interface to be far simpler. 
For the \ehr contest, an even greater portion of break-it scores come from
security bugs.  
This again was by design as we increased the security bug limit. 
Instead of submitting a maximum of two security bugs against a specific build-it team, breakers could submit up to five security (or correctness) bugs against a given team. 
 
Interestingly, making use of advanced analysis techniques did not
factor into the final model; i.e., such techniques did not provide a
meaningful advantage. This makes sense when we consider that such
techniques tend to find generic errors such as crashes, bounds
violations, or null pointer dereferences. Security violations for our
problems are more semantic, e.g., involving incorrect design or
improper use of cryptography. Many correctness bugs were non-generic too, e.g.,
involving incorrect argument processing or mishandling of inconsistent
or incorrect inputs.

Being a build participant and having more coding experience is
identified as a postive factor in the break-it score, according to the
model, but neither is statistically significant (though they are close
to the threshold). Interestingly,
knowledge of C is identified as a strongly negative factor in break-it
score (though again, not statistically significant). Looking closely
at the results, we find that \emph{lack} 
of C knowledge is extremely \emph{uncommon}, but that the handful of
teams in this category did unusually well. However, there are too few
of them for the result to be significant.

\originalmodel{
\begin{table}[t]
\begin{center}
\small
\begin{tabular}{l r r r }
\toprule
\textbf{Factor} & \textbf{Coef.} & \textbf{CI} & \textbf{$p$-value} \\
\midrule
Fall 2015  &  3.847  &  [0.93, 6.76]  &  0.011* \\
\# Team members  &  1.218  &  [0.4, 2.04]  &  0.004* \\
Build participant  &  5.43  &  [1.28, 9.58]  &  0.012* \\
\bottomrule
\end{tabular}
\end{center}
\caption{
ORIGINAL Final linear regression modeling the count of security bugs found by each team. 
Coefficients indicate how many security bugs each factor adds to the
count. 
$R^2 = 0.183$.
}
\label{tab:securitycount-model}
\vspace{-2ex}
\end{table}
}

\begin{table}[t]
\begin{center}
\small
\begin{tabular}{l r r r }
\toprule
\textbf{Factor} & \textbf{Coef.} & \textbf{CI} & \textbf{$p$-value} \\
\midrule
\securelog & --- & --- & ---\ahs{}\\
\significant{\ehr}  &  \significant{9.617}  &  \significant{[5.84, 13.39]}  &  \significant{<0.001*} \\
\significant{\atm}  &  \significant{3.736}  &  \significant{[0.3, 7.18]}  &  \significant{0.035*} \\
\midrule
\significant{\# Team members}  &  \significant{1.196}  &  \significant{[0.35, 2.04]}  &  \significant{0.006*} \\
\midrule
\significant{Build participant}  &  \significant{4.026}  &  \significant{[0.13, 7.92]}  &  \significant{0.045*} \\
%
%
\bottomrule
\end{tabular}
\end{center}
\caption{
Final linear regression modeling the count of security bugs found by each team. 
Coefficients indicate how many security bugs each factor adds to the
count. 
$R^2 = 0.203$.
}
\label{tab:securitycount-model}
\vspace{-4ex}
\end{table}

\paragraph*{Security bugs found}

\begin{figure}[t!]
\begin{center}
\includegraphics[width = 0.85\columnwidth]{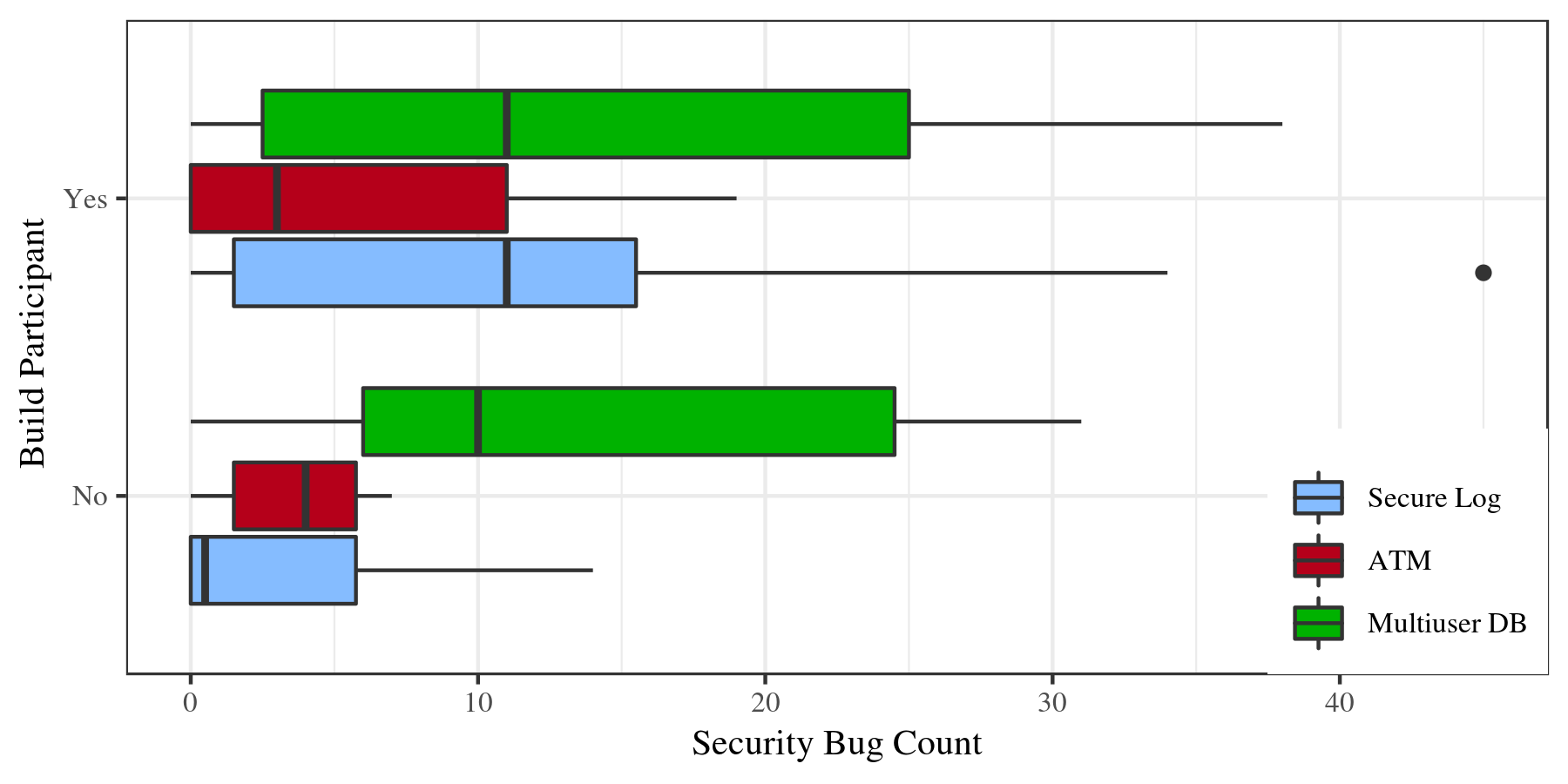}
\end{center}
\caption{Count of security bugs found by each break-it team, organized by contest and 
whether the team also participated in build-it. The heavy vertical
line in the box is the median, the boxes show the first and third quartiles, and the whiskers 
extend to the most outlying data within $\pm1.5\times$ the interquartile range. 
Dots indicate further outliers.
				}
\label{fig:secBugCount-by-builder-box}
\end{figure}

We next consider breaking success as measured by the count of security
bugs a breaking team found.  This model
(Table~\ref{tab:securitycount-model}) 
explains 20\% of variance ($R^2 = 0.203$). 
The model
again shows that team size is
important, with an average of one extra security bug found for each
additional team member. 
Being a qualified builder also significantly helps one's score; this makes intuitive sense, as
one would expect to gain a great deal of insight into how a system
could fail after successfully building a similar system. 
Figure~\ref{fig:secBugCount-by-builder-box} shows the distribution of
the number of security bugs found, per contest, for break-it teams
that were and were not qualified build-it teams. 
%
%
Note that all but eight of the 141 break-it teams made some
attempt, as defined by having made a commit, to participate during the build-it
phase---most of these (115) qualified, but 18 did not. 
%
%
If the reason
was that these teams were less capable programmers, that may imply
that programming ability generally has some correlation with break-it success.

On average, four more security bugs were found by 
\atm teams than \securelog teams. This contrasts with the finding that \securelog
teams had higher overall break-it scores, but corresponds to the finding 
that more \atm submissions had security bugs found against them. 
As discussed above, this is because correctness bugs dominated the \securelog contest 
but were not as dominant in the \atm contest. Once again, the reasons may
have been the smaller budget on per-submission correctness bugs 
for the \atm contest, and the greater potential attack surface in the \atm problem. 

\ehr teams found ten more security bugs on average than \securelog teams.
One possible reason is that the \ehr contest permitted
teams to submit up to five security bug reports per target, rather
than just two. Another is that with \ehr it was easier for breakers to
reuse break tests to see when multiple targets were susceptible to the
same bug.


\subsection{Model differences}
\label{subsec:modeldifferences}

In the conference version of this paper~\cite{Ruef:2016:BBF:2976749.2978382},
we presented previous versions of these models with only data from the \securelog and \atm contests. 
The updated models with \ehr data are very similar to the original
models, but there are some differences. 
We describe the differences to each model in this subsection. 

\paragraph{Ship scores.} 
In the original ship score model, 
students of the MOOC capstone performed 119 points better than non-MOOC teams. 
This correlation goes away when the \ehr data is included.
We hypothesize that this is due to prior coursework. 
MOOC students took three prior security courses that cover cryptography which is directly relevant to
the \securelog and \atm problem, but not the \ehr problem. 

Lines of code is not statistically significant in the updated model, but it was significant in the original model. 
Each additional line of code corresponded with a drop of 0.03 points in ship score. 
Code size may not have improved ship scores for the \ehr contest due to the nature of the problem. 
Teams needed to implement custom access control policies and there are fewer libraries available that implement this functionality. 
%

\paragraph{Presence of security bugs.} 
In the original presence-of-security-bugs model, 
lines of code was a significant factor.
Each additional line of code slightly increased the likelihood of a security bug being present ($1.001\times$). 
Lines of code is not in the latest model, which is similar to the
change in the ship score model (\S\ref{ss:ship-it}). 
We hypothesize this change occured for the same reasons.


\paragraph{Break score.}
The break score model 
basically remained the same
between both versions of the model. 
The only difference is the coefficients slightly changed 
with the addition of 
the \ehr contest. 

\paragraph{Security bugs found.}
The linear regression model for the number of security bugs found
essentially remained unchanged. 
The only material change was the addition of the factor that found that \ehr breakers found more security bugs than \securelog problem breakers. 

\subsection{Summary}

The results of our quantitative analysis provide insights into 
how
different
factors 
%
%
%
correlate with success in building and breaking software. 
Programs written in C and C++ received higher ship scores due to better performance. 
C/C++ submissions were also $11\times$ more likely to
have a reported security flaw 
than submissions written in statically typed languages. 


Break-it teams with more team members found more security bugs and received more break-it points. 
Searching for vulnerabilities is easily parallelizable, so teams with more members 
%
could split the workload and audit more code. 
Successful build participants found more security bugs.
This is intuitive as successfully building a program gives insights into 
the mistakes other similar programs might make. 


\section{Related work}
\label{sec:related}

\bibifi bears similarity to existing programming and security
contests but is unique in its focus on building secure
systems. \bibifi also is related to studies of code and secure
development, but differs in its open-ended contest format.

\paragraph*{Contests} Cybersecurity contests typically
focus on vulnerability discovery and exploitation, and sometimes involve a system
administration component for defense.  
%
One popular style of contest is dubbed \emph{capture the flag} (CTF)
and is exemplified by a contest held at DEFCON~\cite{defcon}. Here,
teams run an identical system that has buggy components. The goal is
to find and exploit the bugs in other competitors' systems while
mitigating the bugs in your own. Compromising a system enables a team
to acquire the system's key and thus ``capture the flag.'' In addition to
DEFCON CTF, there are other CTFs such as iCTF~\cite{Childers2010hacking,Doupe2011live, 205237},
S3~\cite{Antonioli:2017:GIS:3140241.3140253}, KotH~\cite{219742}
and PicoCTF~\cite{picoctf}. The use of this style of contest in an educational setting
has been explored in prior work~\cite{conti2011competition,eagle2013competition,hoffman05exercise}.
The Collegiate Cyber Defense Challenge~\cite{nationalccdc,conklin2006cyber,Conklin2005ccdc} and
the Maryland Cyber Challenge \& Competition~\cite{mdc3} have
contestants defend a system, so their
responsibilities end at the identification and mitigation of
vulnerabilities. These contests focus on bugs in systems as a key
factor of play, but neglect software development. 

Since BIBIFI's inception, additional contests have been developed in its style. 
Make it and Break it~\cite{Yamin:2018:MBI:3284557.3284743} is an evaluation of the Build-it, Break-it, Fix-it type of contest. 
Two teams were tasked with building a secure internet of things (IoT) smart home with functionality including remote control of locks, speakers, and lighting. 
Teams then broke each other's implementations and found vulnerabilities like SQL injection and unauthorized control of locks. 
The contest organizers found this style of contest was beneficial 
in the development of cybersecurity skills and plan to run additional contests in the future. 
Git-based CTF~\cite{219730} is similar to BIBIFI in that students were asked to implement a program according to a given specification. 
It differs in the fact that the CTF was fully run on Github and contestants used issue-tracking to submit breaks. 
In addition, builders were encouraged to fix breaks as soon as breaks were submitted since they periodically lost points for unfixed breaks. 
This seems to have been an effective motivation for convincing builders to fix their mistakes, and it may 
be 
a solution to improve BIBIFI's fix-it participation. 

Programming contests challenge students to build clever, efficient
software, usually with constraints and while under (extreme) time
pressure. The ACM programming contest~\cite{acmprogramming} asks teams
to write several programs in C/C++ or Java during a 5-hour time
period. Google Code Jam~\cite{codejam} sets tasks that must be solved
in minutes, which are then graded according to development speed
(and implicitly, correctness). Topcoder~\cite{topcoder} runs
several contests; the Algorithm competitions are small projects that
take a few hours to a week, whereas Design and Development
competitions are for larger projects that must meet a broader
specification.  Code is judged for correctness (by passing tests),
performance, and sometimes subjectively in terms of code quality or
practicality of design.  All of these resemble the build-it phase of
\bibifi but typically consider smaller tasks; they do not consider the
security of the produced code.

\paragraph*{Secure Development Practices and Advice}
There is a growing literature of recommended practices for secure
development. The BSIMM (``building security in'' maturity
model)~\cite{bsimm} collects information from companies and places it
within a taxonomy. Microsoft's Security Development Lifecycle
(SDL)~\cite{sdl} describes possible strategies for incorporating
security concerns into the development process. Several authors make
recommendations about development lifecycle and coding
practices~\cite{buildsecin,bss,securecode,securecodingC,securecodinganalysis,nistguidelines}.
Acar et al. collect and categorize 19 such resources~\cite{Acar:2017:secdev}.

\paragraph*{Studies of secure software development}
Researchers have considered how to include security in the
development process.
Work by Finifter and
Wagner~\cite{finifter11exploring} and
Prechelt~\cite{prechelt2011plat_forms} relates to both our build-it
and break-it phases: they asked different teams to develop
the same web application using different frameworks, and then
subjected each implementation to automated (black box) testing and
manual review. They found that both forms of review were effective in
different ways, and that framework support for mitigating certain
vulnerabilities improved overall security. Other studies focused on the
effectiveness of vulnerability discovery techniques, e.g., as might be
used during our break-it phase. Edmundson et al.~\cite{codereview}
considered manual code review; Scandariato et
al.~\cite{scandariato2013static} compared different vulnerability
detection tools; other studies looked at software properties that might
co-occur with security problems
\cite{DBLP:conf/issre/WaldenSS14,yang2016improving,
  harrison2010empirical}. 
\bibifi differs from all of these in its
open-ended, contest format: Participants can employ any technique they
like, and with a large enough population and/or measurable impact, the
effectiveness of a given technique will be evident in final outcomes.

Other researchers have examined what factors influence the development of security errors. 
Common findings include developers who do not understand the threat model, security APIs
with confusing options and poorly chosen defaults, and ``temporary"
test configurations that were not corrected prior to deployment~\cite{Fahl:2013bd,Georgiev:2012ht,Egele:2013ei}.
Interview studies with developers suggest that security is often perceived as someone
else's responsibility, not useful for career advancement, and not part of the ``standard''
developer mindset~\cite{Xie:2011tr,Oliveira:2014:PSH:2664243.2664254}. Anecdotal
recommendations resulting from these interviews include mandating and rewarding
secure coding practices, ensuring that secure tools and APIs are more attractive than
less secure ones, enable ``security champions'' with broadly defined roles, and favoring
ongoing dialogue over checklists~\cite{Wurster:2008kd,weir:2017:eusec,becker:2017:eusec}.
Developer Observatory~\cite{205863, 7958576, 205176} is an online platform that enables
large-scale controlled security experiments 
by asking software developers to complete a security relevant programming tasks in the browser. 
Using this platform, Acar et al. studied how developer experience and API design for cryptographic libraries 
impact software security. 
Oliveira et al.~\cite{219422} performed an experiment on security blindspots, which they define as a misconception, misunderstanding, or oversight by the developer in the use of an API. 
Their results indicate that
API blindspots reduce a developer's ability to identity security concerns, I/O functionality is likely to cause blindspots, and experience does not influence a developer's ability to identify blindspots. 
Thompson~\cite{Thompson:2017:LSM:3127005.3127014} analyzed thousands of open source repositories and found that code review of pull requests reduced the number of reported security bugs. 



\paragraph*{Crash de-duplication}
For accurate scoring, BIBIFI identifies duplicate bug reports by
unifying the ones addressed by the same (atomic) fix. But this
approach is manual, and relies on imperfect incentives.
Other works have attempted to automatically de-duplicate bug reports,
notably those involving crashes.
Stack hashing~\cite{Molnar:2009:DTG:1855768.1855773} and
AFL~\cite{afl} coverage profiles offer potential solutions, however Klees et al.~\cite{Klees:2018:EFT:3243734.3243804} show that fuzzers 
are poor at identifying which underlying bugs cause crashing inputs. 
Semantic crash bucketing~\cite{vantonder18sbc} and symbolic analysis~\cite{pham17bucketing} show better results at mapping crashing inputs to unique bugs by taking into account semantic information of the program. 
The former supports BIBIFI's view 
that program fixes correspond to unique bugs.

\section{Conclusions}\label{sec:conclusions}

This paper has presented Build-it, Break-it, Fix-it (\bibifi), a new security
contest that brings together features from typical security contests,
which focus on vulnerability detection and mitigation but not secure
development, and programming contests, which focus on development but
not security.  During the first phase of the contest, teams construct
software they intend to be correct, efficient, and secure. During the
second phase, break-it teams report security vulnerabilities and other
defects in submitted software. In the final, fix-it, phase, builders
fix reported bugs and thereby identify redundant defect reports. Final
scores, following an incentives-conscious scoring system, reward the
best builders and breakers. 

During 2015 and 2016, we ran three contests
involving a total of 156 teams and three different programming
problems. Quantitative analysis from these contests found that the
best performing build-it submissions used C/C++, but submissions coded
in a statically-typed language were less likely to have a security
flaw. Break-it teams that were also successful build-it 
teams were significantly better at finding security bugs. 
Break-it teams with more members were more successful at breaking since auditing code is a task that is easily subdivided.

We have freely released \bibifi at
\url{https://github.com/plum-umd/bibifi-code} to support future research. We believe it can
act as an incubator for ideas to improve secure development. More
information, data, and opportunities to participate are available at
\url{https://builditbreakit.org}.

\paragraph*{Acknowledgments} We thank Jandelyn Plane and Atif Me\-mon
who contributed to the initial development of BIBIFI and its preliminary
data analysis. Many people in the security community, too numerous to
list, contributed ideas and thoughts about BIBIFI during its
development---thank you! Bobby Bhattacharjee and the anonymous
reviewers provided helpful comments on drafts of this paper. This
project was supported with gifts from
Accenture, AT\&T, Booz Allen Hamilton, Galois, Leidos, Patriot Technologies, NCC Group, Trail 
of Bits, Synposis, ASTech Consulting, Cigital, SuprTek, Cyberpoint, and Lockheed
Martin; by a 2016 Google Faculty Research Award; by grants from the NSF under 
awards EDU-1319147 and CNS-1801545; by DARPA under contract FA8750-15-2-0104; and by the U.S. 
Department of Commerce, National Institute for Standards and Technology, 
under Cooperative Agreement 70NANB15H330.

\bibliographystyle{ACM-Reference-Format}
\bibliography{confs_long,proposal,cited,atif,mwh,dml,mlm}

\end{document}